\documentclass[structabstract]{aa}  
\usepackage{graphicx}
\usepackage{txfonts}
\usepackage{longtable}
\usepackage{natbib}

\bibpunct{(}{)}{;}{a}{}{,}

\newcommand{\cgn}{\hbox {CG~12-N }}
\newcommand{\cgnp}{\hbox {CG~12-N}}
\newcommand{\cgs}{\hbox {CG~12-S }}
\newcommand{\cgsp}{\hbox {CG~12-S}}

\newcommand{\cgswp}{\hbox {CG~12-SW}}

\newcommand{\twco}{{\hbox {\ensuremath{\mathrm{^{12}CO}} }}}
\newcommand{\ceo}{{\hbox {\ensuremath{\mathrm{C^{18}O}} }}}

\newcommand{\dcopl}{{\hbox {\ensuremath{\mathrm{DCO^+}}}\ }}
\newcommand{\dcoplp}{{\hbox {\ensuremath{\mathrm{DCO^+}}}}}

\newcommand{\Msun}{\ensuremath{\mathrm{M}_\odot}}

\newcommand{\kkms}{\ensuremath{\mathrm{K\,km\,s^{-1}}}}

\newcommand{\mum}{\mu {\rm m}}
\def\Ks{\hbox{$K$s}}
\def\K{\hbox{$K$}}
\def\Js{\hbox{$J$s}}
\def\J{\hbox{$J$}}
\def\H{\hbox{$H$}}
\newcommand{\umag}{{$^{m}$}}
\newcommand{\ammonia}{NH{$_3$}} 
\def\jhks{\hbox{$J\!H\!K$s}}              
\def\JHKL{\hbox{$J\!H\!K\!L$}}              
\def\JHKLM{\hbox{$J\!H\!K\!L\!M$}}              
\def\fm{\hbox{$.\!\!^{\rm m}$}}
\newcommand{\CCD}{\J-\H/\H-\Ks \ diagram}

\newcommand{\one}{58}

\newcommand{\two}{74}

\newcommand{\three}{86}

\newcommand{\four}{87}

\newcommand{\five}{108}

\newcommand{\six}{120}

\newcommand{\nnn}{131}

\newcommand{\wsixa}{132}

\newcommand{\wsixb}{138}

\newcommand{\fourone}{41}
\newcommand{\eightone}{83}
\newcommand{\twooei}{216}

\begin{document}
   \title{Near infrared imaging of the cometary globule CG~12 \thanks{Based 
on observations collected at the European Southern Observatory,  La Silla, Chile} \thanks{Table 1 is only available in electronic form
at the CDS via anonymous ftp to cdsarc.u-strasbg.fr}}

   \author{L\,K.  Haikala \inst{1}       
      \and
          B. Reipurth\inst{2}
         }

 \institute {Observatory, PO Box 14, University of Helsinki, Finland\\
              \email{lauri.haikala@helsinki.fi}
     \and
              Institute for Astronomy, University of Hawaii, 
640 N. Aohoku Place, Hilo, HI 96720, USA\\
              \email{reipurth@ifa.hawaii.edu}}

   \date{}

 
  \abstract
  {Cometary globule 12 is a relatively little investigated medium- and
    low mass star forming region 210 pc above the Galactic plane.}
   {This study sets out to discover the possibly embedded members of
     the CG~12 stellar cluster, to refine the NIR photometry of the
     known member stars and to study the star formation activity in
     CG~12 and its relation to the distribution of molecular gas, dust
     and mid- to far-infrared emission in the cloud.}
   { NIR \J, \H, and \Ks \ imaging and stellar photometry is used to
     analyse the stellar content and the structure of CG~12.}
{Several new members and member candidates of the CG~12 stellar
  cluster were found. The new members include in particular a highly
  embedded source with a circumstellar disk or shell and a variable star
  with a circumstellar disk which forms a binary with a previously
  known A spectral type cluster member.  The central source of the
  known collimated molecular outflow in CG~12 and an associated
  ``hourglass''-shaped object due to reflected light from the source
  were also detected.  The maximum visual extinction in the
  cloud, based on observations of background stars, is $\sim$20\umag, but
  this is only a lower limit for the extinction through the two dense
  cloud cores.
HIRES-enhanced IRAS images are used together with SOFI \jhks \ imaging
to study the two associated IRAS point sources, 13546--3941 and
13547--3944. Two new 12\,$\mum$ \ sources coinciding with NIR excess
stars were detected in the direction of IRAS 13546--3941. The IRAS
13547--3944 emission at 12 and 25\,$\mum$\ originates in  the Herbig~AeBe
star h4636n and the 60 and 100\,$\mum$\ emission from an adjacent cold
source.}
   {}


   \maketitle
%
\titlerunning{NIR imaging of the cometary globule CG~12}
\authorrunning{L\,K. Haikala \& B.Reipurth}
\section{Introduction} \label{sec:introduction} 

Low and intermediate mass star formation takes place primarily in
isolated clusters (e.g., the CrA star forming cloud) and to a lesser
extent in low mass star forming regions (e.g., Taurus and
Chamaeleon).  As the stars form in dense gas/dust clouds, which are
tightly concentrated towards the Galactic plane, star formation far out
of the plane is not a likely event. Even though Chamaeleon and CrA are
high Galactic latitude regions, they lie well within the scale height
of molecular interstellar matter in the Galaxy (50-75 pc).  The well
known nearby star forming regions have been studied in great detail
from visual to radio wavelengths, but less is known of star formation
regions off the Galactic plane.

\citet{VanTilletal1975} detected strong CO emission in the direction
of the reflection nebulosity \object{NGC 5367} and, assuming a
distance of 300 pc, calculated a mass of 30 \Msun \ for the mapped
area.  NGC 5367 lies in the head of the Cometary Globule 12,
\object{CG 12}, listed in \citet{HawardenBrand1976}.  Despite being
classified by \citet{HawardenBrand1976} as a cometary globule,
together with those in the Gum nebula \citep{reipurth1983}, 
CG 12 is actually a high-latitude 
low- and intermediate-mass star formation region.  With a
Galactic latitude of 21\degr \ and at the distance of $\sim$550\,pc
\defcitealias{getmanetal2008}{GFLBG} (\citet{maheswar1996} and
\citet{getmanetal2008}, hereafter GFLBG), CG~12 lies more than 200~pc
above the plane.  A loose stellar cluster including at least three
late B stars and two A stars \citep{williams1977} is associated with
CG~12. The associated reflection nebula  NGC 5367 is illuminated by
the binary star \object{h4636} (spectral types B4V and B7V). This
binary contains at least one, or possibly two, Herbig~AeBe stars
\citep{ReipurthZinnecker1993}.  \citetalias{getmanetal2008} find a
high concentration of X-ray sources in the CG\~12 region mainly in the direction of visible
or NIR stellar objects and conclude that more than 50 of these are
T-Tauri stars associated with the nebula.  \citet{white1993} mapped
the CG~12 centre region in CO (2--1) and \ceo (2--1) and found a compact core
and a highly collimated molecular outflow. Further molecular line
observations of CG~12 are presented in \citet{yonekuraetal1999a} and
recently in\defcitealias{haikalaolberg2007}{Paper 1}
\citet{haikalaolberg2007} \citepalias{haikalaolberg2007} 
and\defcitealias{haikalaetal2006}{Paper 2} \citet{haikalaetal2006}
\citepalias{haikalaetal2006}. 
The structure and dynamics of the CG~12
core have been discussed in detail in Papers 1 and 2.  The molecular
material lies in a 10$'$ NS elongated lane with three compact \ceo
maxima, \cgnp, \cgsp, and \cgswp. The southern maximum, \cgsp, lies
near the binary h4636.  The \ceo emission in \cgs traces mainly warm
gas on the surface of a dense core (called \dcoplp \ core) revealed in
\dcopl and CS emission and in 1.2\,mm continuum
\citepalias{haikalaolberg2007, haikalaetal2006}. The centre of the
highly collimated molecular outflow reported in \citet{white1993} lies
in the direction of the \dcopl core.  The northern core, \cgnp, is
cold and the relative velocities and intensities of \ceo and high
density tracers indicate that molecular material is highly depleted.
The molecular mass of the CG~12 core area, as traced by the \ceo
observations, is larger than 100 \Msun. This mass corresponds only to
the area observed in \ceo and traced by the \ceo emission. The mass of
the cloud envelope not visible in \ceo and that outside the mapped
area is much larger, as shown by e.g. the large scale CO mapping of
\citet{yonekuraetal1999a} according to which the mass of the globule
is larger than 300 \Msun.  The linear size of the CG~12 core region ($>$1 pc) 
and the molecular
mass are of the same order as those of known nearby low mass
star formation regions like Chamaeleon~I.  CG~12 has been recently
reviewed by \citet{2008hsf2.book..847R}.

The median photometric age of the T-Tauri population in CG~12 is
$\sim$4 Myr, but the apparent spread of ages is considerable, from 1
to 20 Myr \citepalias{getmanetal2008}.  The more than 50 T-Tauri stars
detected by \citetalias {getmanetal2008} represent possibly the first
generation of stars born in the CG~12 area, and the B and A stars
\citep{williams1977} may be the second generation.  The four IRAS
point sources and the collimated molecular outflow in the CG~12 cloud
show that star formation is presently going on.  Because of the great 
distance and the high extinction towards the cloud core, the limiting
magnitudes of the Two Micron All Sky Survey (2MASS) survey allow only
the detection  of the brightest stars embedded in the cloud in all  three
\jhks \ colours.  A deeper \J,\H,\Ks \ survey of CG~12 than 2MASS is
clearly needed to study the possible embedded stellar population.
Spitzer NIR IRAC imaging data on CG~12 has recently become public and
can also be used to analyse the embedded stars.

We report imaging of the core of CG~12 in \J, \H,  and \Ks \ NIR bands with
SOFI at the NTT telescope at La Silla.
Observations, data reduction, and calibration procedures are described
in Sect.~\ref{sec:observations} and the observational results in
Sect~\ref{sec:results}. The new results are discussed and compared
with available data at other wavelengths in
Sect.~\ref{sec:discussion}.  The results are summarised and
the conclusions drawn in Sect.~\ref{sec:summary}.


\section{Observations and reductions}  \label{sec:observations} 

\begin{figure*}
\centering
\includegraphics [width=16cm] {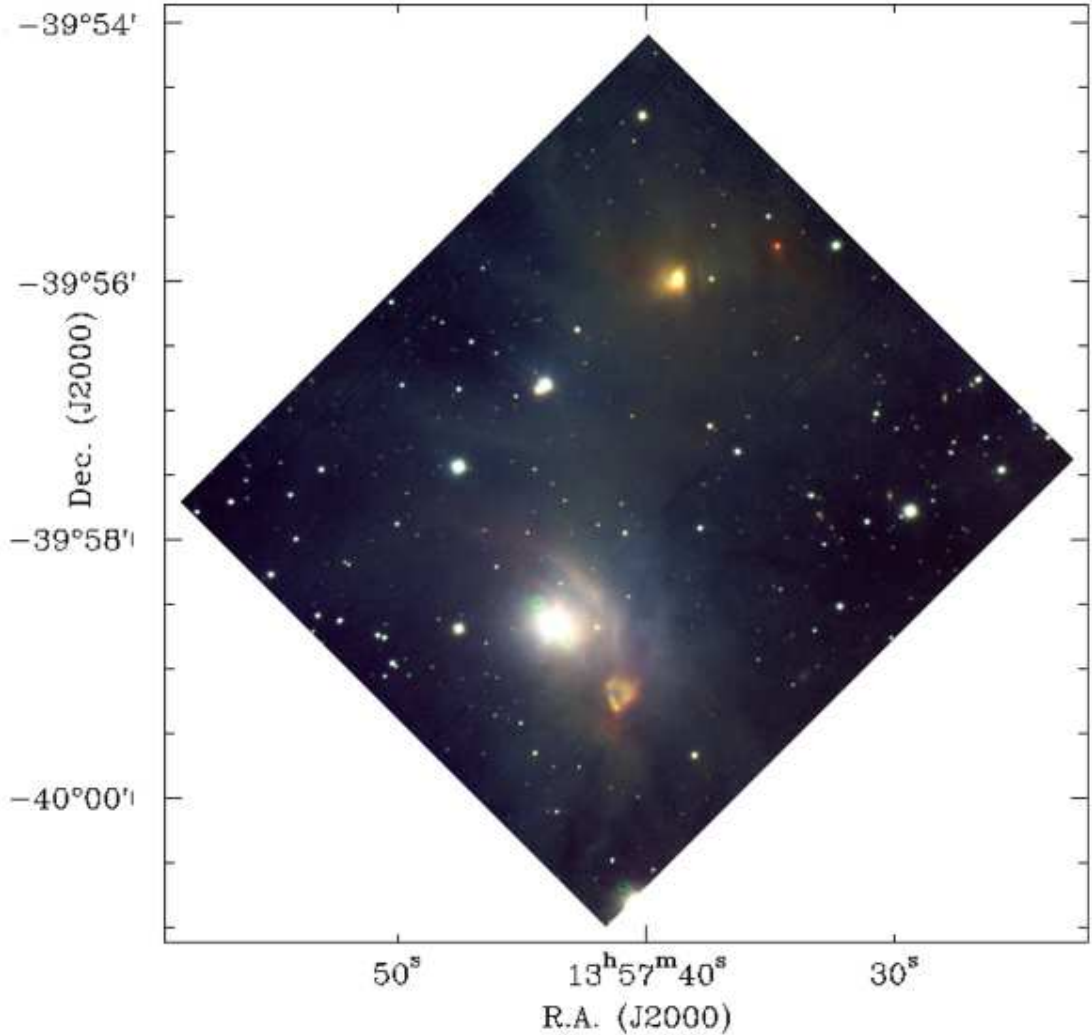}
   \caption{Colour coded SOFI image of CG~12. The \J, \H, and \Ks \ bands
     are coded in blue, green and red, respectively. Square root
     scaling has been used to better bring out the faint surface
     brightness structures.}
\label{figure:CG12_3col}
\end{figure*}

CG~12 was  imaged in \J, \H, and \Ks \ with SOFI on the NTT during two
observing runs, May 20-24, 2005 and June 13-15, 2007. In 2007 the \Js
\ filter was used instead of the \J.  In order to optimise the
coverage in the north-south direction the detector array was rotated
45 degrees. Because of the high surface brightness in CG 12 the
observations were carried out in the on-off mode instead of the
standard jitter mode. After each on-integration an off-position
outside the globule was observed and jittering was done after each
on-off pair. The 2005 \H \ and \Ks \ integrations consisted of 10
individual 6 second integrations.  The 2005 \J \ and all the 2007
observations had a 10 second basic integration time.  The average seeing
was $\sim$0\farcs8 \ and $\sim$0\farcs7 \ in 2005 and 2007,
respectively. The observed on-source times were 40, 24, and 30 minutes
in 2005 and 65, 24, and 41 minutes in 2007 in the \J, \H, and \Ks,
respectively. Standard stars from the \citet {perssonetal1998} faint
NIR standard list were observed frequently during the nights.

The IRAF{\footnote {IRAF is distributed by the National Optical
Astronomy Observatories, which are operated by the Association of
Universities for Research in Astronomy, Inc., under cooperative
agreement with the National Science Foundations}} external XDIMSUM
package was used in the data reduction.  The images were searched for
cosmic rays, sky subtracted, flat fielded, illumination corrected,
registered and averaged.  The two off-position images nearest in time
to each on-position image were used in the sky-subtraction. An object
mask was constructed for each off-position image. Applying these masks
in the sky-subtraction produced hole-masks for each sky-subtracted
image.  Special dome flats and illumination correction frames provided
by the NTT team were used to flat-field and to illumination-correct
the sky-subtracted images. Rejection masks combined from a bad pixel
mask and individual cosmic ray and hole masks were used when averaging
the registered images.

A bright star in an image produces an artifact (inter-quadrant row
cross talk) which appears as a stripe at the same line as the star and
symmetrically on the other half of the detector. This can be corrected
if the star in question is not saturated (see the SOFI
manual). However, in CG~12 both the components of the binary h4636 are
heavily saturated and the striping cannot be corrected.  Reflections
in the \H \ filter cause ring-like ghosts around bright stars.  No
effort was made to correct this.

The intensity scale was monitored using the standard star observations
before and after each observing block.  The 2007 \Js \ counts were 5\%
smaller than the 2005 \J \ counts. The 2007 \Js \ data were scaled
to bring the data sets into the same scale. The possible slight 
colour-dependent difference due to the narrower \Js \ filter was ignored.

\addtocounter{table}{1}

The SExtractor software v. 2.5.0 \citep {bertinarnouts1996} package
was used to detect the stars and galaxies in the images. 
Altogether 423 sources were detected in all the three filters. The
galaxies were excluded both by using the SExtractor star keyword and by visual
inspection of the images. Strong surface brightness filaments also
appeared as extended and elongated sources in the photometry.  After
elimination of sources estimated to be galaxies and sources 
due to surface brightness, 280 sources remained. Of these 218  have
formal errors less than 0\fm15  in all colours. If only \H \ and \Ks \ 
filters are considered the number amounts to 227.  The magnitude
zero points of the added-up data were fixed using the standard star
measurements from the 2005 observing run. The instrumental magnitudes
were first converted to the \citet {perssonetal1998} photometric
system and then to the 2MASS photometric system as described in \citet
{ascensoetal2007}. Finally, the magnitude scale was checked by
comparing the SOFI photometry of stars in common with 2MASS and which
had good quality 2MASS photometry.  Apart from the photometry zero point
check the transformation to the 2MASS photometric system was done to
make the comparison with literature data easier.  The limiting magnitudes
(for a formal error of 0\fm15) are approximately 21\fm0, 20\fm5 and 20\fm0
for \J, \H,  and \Ks, respectively.  The limiting magnitude, however, varies
over the observed area and is less in the
regions of strong surface brightness and also where the brightness gradient
is strong. This concerns especially the surroundings of the
binary h4636. 

The components of the binary h4636 were strongly saturated and no
photometry was possible.  The counts of some of the brightest stars in
the field were within the saturation limit. The \H \ and \Ks \ photometry
for these stars was obtained using the 2005 observations with a 6
second basic integration time. Even if the stars were not saturated
their count levels were in the nonlinear region of the detector.  An
extreme nonlinearity of 5\% would produce an error of 0\fm05.  This
will be noted below where the photometry is discussed.  The photometry
of the two close binaries, the \citet{williams1977} star W77-6
and the one in the direction of \cgnp, is challenging because of the
brightness of the stars and the strong background nebulosity. These
stars were measured manually by varying both the source and background
apertures to reach a satisfactory compromise. As a check, the chosen
aperture settings were also applied to single bright stars, the results 
of which were compared with those  from the automated SExtractor
photometry. Even though the single stars agreed well, one should bear
in mind that the uncertainty in the binary photometry exceeds the
formal error of the measurement.

The results of the SOFI stellar photometry are tabulated ordered by
increasing right ascension in the online Table
\ref{table:jhkphotometry}.  Columns one to three list our stellar
sequence number, other ID, and the J(2000) coordinates,
respectively. The \J, \H, and \Ks \ magnitudes with their formal
errors and the (\J-\H), (\H-\Ks) colours are listed in columns four to
eight, respectively.  Comments are given in column nine. The stars
considered as members or potential members of the CG\,12 stellar
cluster (see Sects. \ref{sec:star1} to \ref{potentialmembers})
are identified in the comments column with 1 and 2, respectively. The
\citetalias{getmanetal2008} Chandra X-ray source designation I\_NN,
where I stands for the \citetalias{getmanetal2008} field I and NN the
sequence number, is also given for stars that are the same as in that
publication. For the identification with the Chandra objects a
  positional coincidence better than 1\arcsec \ was required for stars
  not already identified with 2MASS stars by
  \citetalias{getmanetal2008}.

\section{Results} \label{sec:results} 

\begin{figure} \centering \includegraphics 
   [width=8cm, angle=-90.0]{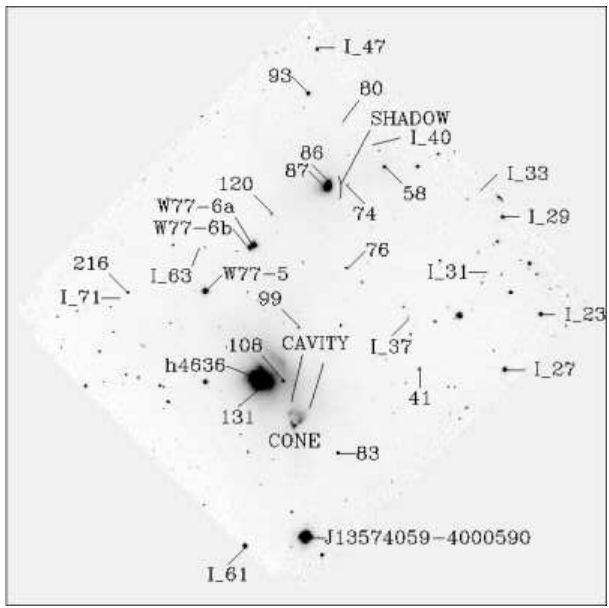}
   \caption{ SOFI \Ks \ band image. Notable features in the cloud are
     identified. The stars in the online Table
     \ref{table:jhkphotometry} are identified with their sequence
     number. W77-5 and -6 refer to the \citet{williams1977}
     identification.  The 2MASS J13574059-4000590 and the
     \citetalias{getmanetal2008} objects (I\_NN) not in Table
     \ref{table:jhkphotometry} are also indicated.}
\label{figure:features}
\end{figure}

\subsection{Imaging} \label{sec:imaging} 

 The on-off observing method used in this study preserves also the
extended surface brightness which is smeared out if the standard jitter
observing method is used. Only the surface brightness gradients are
visible in the final reduced images if the simple jitter method is
used.  Extended surface brightness due to scattered light is seen in
all three colours. However, it is not possible to determine the
absolute level of this surface brightness. Therefore, the surface
brightness level was arbitrarily set to zero in a small region in the
western part of the images where visual inspection shows
that the surface brightness is the smallest. However, the real
background zero level, even in the \Ks \ band where it is the lowest,
most probably lies outside the imaged area.

A false colour SOFI image of CG~12 is shown in
Fig.~\ref{figure:CG12_3col}. The \J, \H,  and \Ks \ images are coded
in blue, green, and red respectively. The image is dominated by the
bright binary h4636 and the reflection nebulosity it is illuminating.
The reflection nebulosity surrounding h4636 is predominantly bluish in
colour whereas the northern part of the image is reddish.  The number
of stars (and galaxies) in the western and eastern parts of the image
is high compared to the north-south diagonal where at some locations
hardly any stars are seen. The number density of stars anticorrelates,
as expected, with the observed \ceo emission in the globule
\citepalias{haikalaolberg2007}.  Other noteworthy features are the
cone shaped object to the SW of h4636, a nebulous binary in the
northern part of the globule, and a dark lane to the West of it.  A
very red object (star \one) lies west of this binary near the
edge of the image.  An elongated cavity, best visible in the \J
\ image, is seen north of the cone. The location of these
features in the SOFI image and selected stellar objects to be
discussed below are shown in Fig. \ref{figure:features}.  The
positions of these objects relative to the IRAS point sources, the
\ceo cores (\cgs and \cgnp), the \dcopl emission and the \twco
molecular outflow \citep{white1993} are shown in Fig. 10 in Paper
1. The cone is projected on the \dcopl core and the centre of the
collimated molecular outflow \citep{white1993} but it lies 
south of the \ceo maximum.  The nebulous binary in the North lies at
the edge of \cgnp.

\subsection{Photometry} \label{sec:photometry} 

\begin{figure} \centering \includegraphics 
[width=8cm]{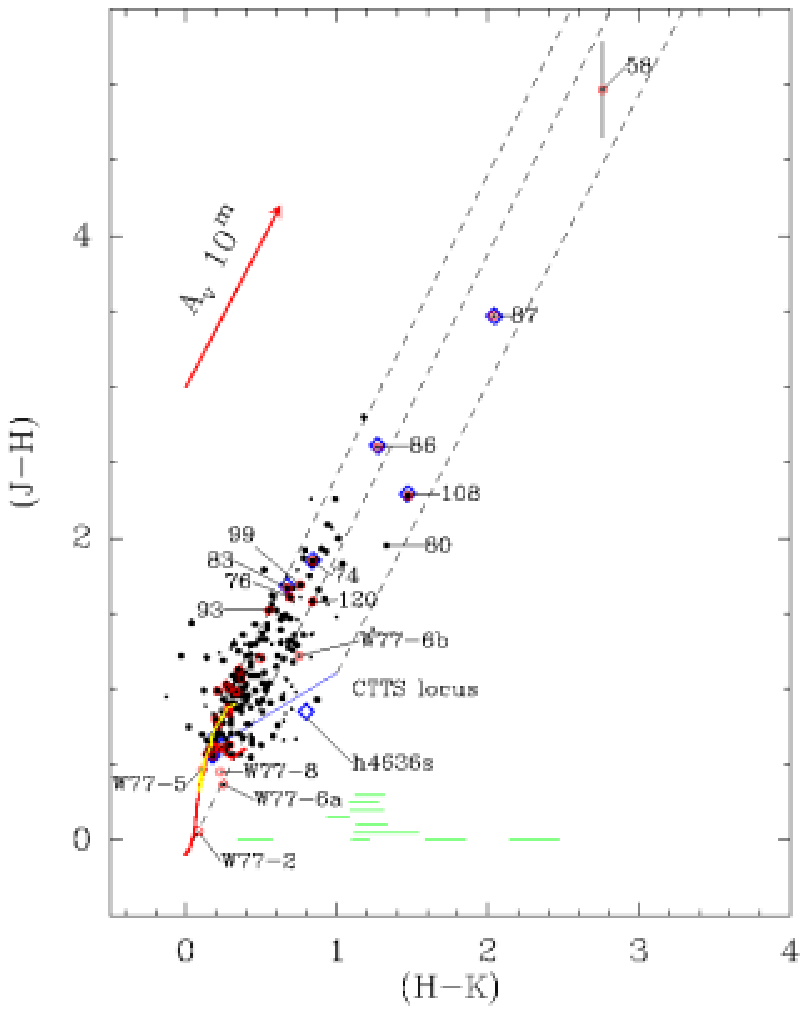}
\caption{ \CCD \ for stars with formal \jhks
  \ errors less than 0\fm15. The stars with (\J-\H) and/or
  (\H-\Ks) error larger than 0\fm1 are plotted using a small
  symbol. The stars for which the \J \ error is large but which are
  well detected in \H \ and \Ks \ are indicated with horizontal green
  (\H-\Ks) error bars at the bottom of the plot.  The main-sequence,
  the giant branch and the M0 CTTS-locus are plotted in red, yellow and
  blue respectively. Blue diamonds indicate stars detected in X-ray
  \citepalias{getmanetal2008}. Red circles identify the stars which
  are brighter than 15\fm0 in \Ks. The
  dashed lines show the effect of visual extinction in the diagram
  according to the \citet{besselbrett1988} reddening law, with the red arrow
  corresponding to Av of 10\umag.}
\label{figure:jhk}
\end{figure}

The \J-\H/\H-\Ks \ colour-colour diagram for stars with \J, \H, and
\Ks \ formal errors less than 0\fm15 is shown in
Fig.~\ref{figure:jhk}. The only star in the image with a larger error
in \J \ is the highly reddened star \one.  The main-sequence, the
giant branch \citep{besselbrett1988} and the M0 CTTS-locus
\citep{meyeretal1997}, all transformed to the 2MASS photometric
system, are plotted in red, yellow, and blue respectively. The
  dashed lines show the effect of visual extinction in the diagram
  according to the \citet{besselbrett1988} reddening law, with  the red arrow
  corresponding to Av of 10\umag. It is assumed in the figure that
  the slope of the reddening line is constant even for high
  extinctions. A small symbol is used if the (\J-\H) and/or the
(\H-\Ks) formal error is larger than 0\fm1. The stars with \Ks
\ magnitudes 15 or less are indicated with red circles.  The blue
diamonds indicate X-ray detections \citepalias{getmanetal2008}
associated with stellar objects in the SOFI image.  The \CCD
\ includes the \citet{williams1977} stars 2 and 8 and the star 2MASS
J13574059-4000590, which are not inside
Fig. \ref{figure:CG12_3col}. The colours for these stars, the
\citet{williams1977} star 5 and for h4636s, which were saturated in
the SOFI images, were taken from the 2MASS catalog.  The (\H-\Ks)
index of stars which had reliable photometry only in the \H \ and \Ks
\ bands are indicated with green error bars at the bottom of Fig.
\ref{figure:jhk}.

Even though the obvious extragalactic sources were filtered out from
the original data set, it is highly likely that some of these sources
still remain in the final data set. Likely candidates are e.g. the
faint outliers below the CTTS locus and to the right of the reddening
line and objects with negative (\H-\Ks) colour.

\section{Discussion} \label{sec:discussion} 

 The new NIR imaging and photometry of the CG~12 head, together with
 data available at other wavelengths, is used to analyse the
 associated embedded stellar population. The data additional to the
 present NIR observations are in particular from the 2MASS survey, the
 data from the IRAS satellite, the Chandra X-ray data
 \citepalias{getmanetal2008}, and Spitzer NIR imaging. In the
 following the NIR photometry is discussed first, then the extended
 surface emission and finally, the IRAS observations.

\subsection{Photometry} \label{sec:discphotometry}

 The distance to CG~12 is uncertain.  \citet{maheswar1996} and
 \citetalias{getmanetal2008}, using different methods, suggest a
 distance of 550\,pc.  This is less than the 630\,pc obtained by
 \citet{williams1977} using main sequence fitting.  The distance
 modulus ranges therefore between 8\fm6 and 9\fm0.  The strong
 extinction in the direction of CG~12 starts at the distance of CG~12
 \citep{maheswar1996}.  One would therefore expect to see not only
 stars embedded in the cloud as well as background stars but also unreddened
 foreground objects. The number of these foreground stars can be
 estimated as follows. \citet {mattila1980} gives the  number of
   main sequence stars from late O to M spectral types and giants from
   F8 to M8 as a function of spectral type and Galactic latitude in
   the local Milky Way. The estimated number of CG~12 foreground stars
   in a 5\arcmin \ by 5\arcmin \ field, at a Galactic latitude of
   21\degr \ is $\sim$6. 
   The expected foreground stars are of spectral type G5 or later and
   none earlier. The number of giant stars in the field (including
   also the background) is predicted to be smaller than 1.  As argued
   by \citetalias{getmanetal2008}, the number of the known B and A
   type stars associated with CG~12 predicts a population of $\sim$80
   stars with masses 0.5\,\Msun$<$ M $< 2$\, \Msun. The SOFI image
   should therefore contain a considerable fraction of this unknown
   population, and if not too deeply embedded, they should be visible
   in the image. The \citet{mattila1980} model does not predict any
   early type field stars other than by chance at this high Galactic
   latitude region. The early type stars seen in CG~12 are therefore
   likely to be members of the CG~12 stellar population.

The unreddened foreground stars are likely to be among the
 brightest stars in the SOFI field.  The absolute \K \ magnitude of
 late main sequence  M stars is less or
   equal to 5\umag \ \citep{besselbrett1988}. An unreddened G0 or M5
   main-sequence star at a distance of 550\,pc has a \K \ magnitude of
   11\fm6 (12\fm0 in \J) and 14\fm7 (15\fm7 in \J) respectively.
   This indicates that  stars fainter than
   15\fm0 in \K \ are either embedded in CG~12, or are background
   objects.  Stars brighter than this can be either foreground
   objects, embedded in CG~12 or background stars.  The stars brighter
   than 15\umag \ in \K \ in the SOFI field are indicated by red
   circles in Fig. \ref{figure:jhk}.  As predicted by the Galactic
   stellar model some of these stars have non-reddened late
   main-sequence colours. A larger group of these stars have colours
   indicating extinction less than 5\umag \ in the visual (assuming
   they are not early-type main-sequence stars). Many of these could
   be background F-G type stars.  A group of bright stars is located
   around (\J-\H) 1\fm7, (\H-\Ks) 0\fm8 in Fig. \ref{figure:jhk}
   indicating a visual extinction of 10\umag \ or larger. Stars \two
   \ and \six \ are among these stars.  It is not possible from the
   \jhks \ photometry alone to decide if these stars are embedded or
   background stars.  Ten stars in the direction of the CG~12 cloud
   have reliable photometry only in \H \ and \Ks \ bands (error $<$
   0\fm15 in both bands). The (\H-\Ks) colours of these stars are
   indicated by the green (\H-\Ks) error bars in the lower part of
   Fig. \ref{figure:jhk}. With the exception of one object the visual
   extinction towards these stars is likely to be 15\umag \ or more.

\begin{table*}
\caption{\jhks \  photometry for selected stars in CG~12} \label{table:jhkphot}      
\label{table:1}      
\centering    
\begin{tabular}{l c c c c c c}     
\hline\hline       
          Star    &        \J       &               \H     &             \Ks          &  (\J-\H)  &  (\H-\Ks)  & Comment\\
                  &        mag      &              mag     &             mag          &   mag     &  mag       & \\
\hline       
         58    &    21.62 $\pm$ 0.32   &    16.65  $\pm$ 0.01   &   13.89 $\pm$ 0.01    &   4.97  &   2.76   &\\
         74    &    17.62 $\pm$ 0.01   &    15.77  $\pm$ 0.01   &   14.93 $\pm$ 0.01    &   1.85  &   0.84   &\\
         86    &    15.39 $\pm$ 0.01   &    12.67  $\pm$ 0.01   &   11.39 $\pm$ 0.01    &   2.72  &   1.28   &\\
         87    &    16.63 $\pm$ 0.01   &    12.95  $\pm$ 0.01   &   10.81 $\pm$ 0.01    &   3.68  &   2.14   &\\
        108    &    18.13 $\pm$ 0.09   &    15.85  $\pm$ 0.05   &   14.38 $\pm$ 0.01    &   2.29  &   1.47   &\\
        120    &    16.59 $\pm$ 0.02   &    15.02  $\pm$ 0.01   &   14.18 $\pm$ 0.01    &   1.57  &   0.85   &\\
        131    &    20.27:             &    17.98  $\pm$ 0.06   &   16.69 $\pm$ 0.03    &   2.30  &   1.29   &\\
     h4636s    &     8.87 $\pm$ 0.01   &     8.07  $\pm$ 0.09   &    7.26 $\pm$ 0.06    &   0.80  &   0.81   &2MASS\\
       W77-5    &    11.19 $\pm$ 0.01   &    10.73  $\pm$ 0.01   &   10.69 $\pm$ 0.01    &   0.46  &   0.04   &2MASS\\
      W77-6a    &    11.37 $\pm$ 0.10   &    10.95  $\pm$ 0.10   &   10.72 $\pm$ 0.10    &   0.42  &   0.23   &\\
      W77-6b    &    13.93 $\pm$ 0.10   &    12.79  $\pm$ 0.10   &   12.03 $\pm$ 0.10    &   1.14  &   0.76   &\\

\hline                  
\end{tabular}
\end{table*}

\subsubsection{Star \one \ (\object{J 135734.76-395543.8})} \label{sec:star1}

This is the reddest object in the cloud. The star is only marginally
detected in \J \ (21\fm62$\pm$0\fm32) but is bright in \H \ and \Ks \
(16\fm61 and 13\fm89, respectively). Its position in the \CCD \
indicates IR excess and reddening corresponding to an Av of 35\umag \
or more.  Because of the large error in the (\J-\H) index the star   
could also be a highly reddened giant.  If the star were a background giant
it would, however, be difficult
to explain the origin of the reddening. The \ceo (1-0) integrated line
intensity in the direction of this star is 2.5 \kkms \ in the main
beam temperature scale (it is significantly offset from the \ceo column
density maximum, \cgn \citepalias{haikalaolberg2007}) corresponding to
20\umag \ of Av.  This is only half of the extinction deduced from the
\CCD.  There is also no hint of a dense core where CO could be depleted
in the SIMBA 1.2 mm continuum map (Fig.3, Paper 2). By chance the
position in CG 12 where \citet{bourkeetal1995a} observed \ammonia \ is
only 20$''$ north of the position of this star and therefore well within
the 1\farcm4 beam.  The observed 0.11K \ammonia \ antenna
temperature does not indicate the presence of a localised dense core
at this position.

One should, however, be very careful when assuming a one to one
correspondence of  the calculated \ceo column density and optical
extinction because the relation is statistical.  The calculated \ceo
column density depends strongly on the assumed LTE condition, the
adopted excitation temperature, the assumed beam filling factor
and the molecular abundance used in the calculations
\citepalias{haikalaolberg2007}. The \ceo maximum \cgs highlights this
problem. The \ceo emission does not trace the maximum column density,
which lies in the direction of the \dcopl core, but warm gas on its
surface \citepalias{haikalaolberg2007,haikalaetal2006}. The molecular
line observations of optically thin transitions like \ceo trace the
total cloud column density.  There is, however, no justification for the
assumption that a star embedded in the cloud and seen in the direction of
a molecular column density maximum necessarily lies behind the
maximum.  The extinction towards a star corresponds to a pencil beam
whereas the \ceo column densities (and  column densities of other
molecules) are calculated from observations made with a gaussian beam
and represent therefore an average over the beam. The half power beam
widths of the CG~12 \ceo observations presented in Paper 1 are 47$''$
and 24$''$ for the (1-0) and (2-1) transitions respectively.  At the
assumed distance of CG 12, 550\,pc, 47$''$ corresponds to 0.125\,pc. 

The reddish surface brightness surrounding star \one \ in
Figs. \ref{figure:CG12_3col} and \ref{figure:KsHIRESnorth} indicates
that the star is embedded in the CG~12 cloud. Therefore, a plausible
explanation for the high reddening is that the star is embedded in the
CG2 cloud and the reddening takes place in a circumstellar disk. Such
a disk does not show up in single dish molecular line observations
because of severe beam dilution.  Star \one \ is not detected in the
Chandra observations \citepalias{getmanetal2008} but is visible in the
Spitzer IRAC images in all four bands from 3.6 $\mum$ to 8 $\mum$. The
star is a Class~I candidate embedded in the CG~12 cloud.

\subsubsection{Stars \two,\ \three,\ \four \ and \six} \label{sec:stars234}

Stars \two\ (\object{J 135737.49-395559.5}), \three\
(\object{J 135738.92-395558.7}), and \four \ (\object{J 135738.95-395600.7})
were detected in X-rays (GFLBG objects I\_41, I\_45, and I\_46,
respectively).  Photometry for these stars is summarised in Table
\ref{table:jhkphot} where the SOFI/2MASS  \jhks \ photometry for stars 
discussed in detail in the text is listed.
The position of star \two \ in the \CCD \ shows that the visual extinction
towards this star is $\sim$10 magnitudes. It has no surrounding
nebulosity, but its X-ray activity \citepalias{getmanetal2008}
indicates that it is likely to be associated with CG~12.  In the
Spitzer images it is visible only at the two shortest wavelengths.

Stars \three \ and \four \ form a nebulous binary in the direction of
the \ceo maximum \cgn. The approximate visual extinction towards stars
\three \ and \four \ is larger than 15 and 20 magnitudes
respectively. This is significantly more than the 8-10 magnitudes
estimated by \citetalias{getmanetal2008} from their X-ray
spectrum. Star \four \ has NIR excess in 
  Fig. (\ref{figure:jhk}).  As discussed in Sect.
\ref{sec:observations} the photometry of these stars is difficult
because of the surrounding background nebulosity and the brightness of
star \four \ in the \Ks \ band. A further uncertainty in
  assessing the nature of star \four \ is the uncertainty of the slope
  of the reddening curve and its colour dependence.
However, the high extinction towards the two stars and the
NIR excess of star \four \ are not in doubt.  In the Spitzer images
the stars \three \ and \four \ are surrounded by a nebulosity. Both
stars are saturated in all but the 8 $\mum$ image.

The location of star \six \ (\object{J135743.10-395624.0}) in the \CCD
\ indicates a possible small infrared excess and approximately a visual 
extinction Av of 7\umag \ if originating from the CTTS locus.  The
star is visible in all the Spitzer NIR images.  There is no indication
of nebulosity around the star in the \jhks \ nor in the Spitzer images
and it was not detected by Chandra \citepalias{getmanetal2008}. The
de-reddened magnitudes suggest that the star is associated with CG~12.

\subsubsection{ h4636, stars \five \ and  \nnn.}\label{sec:starsHJ5nn} 

The components of the binary h4636 are the brightest objects in
the SOFI image. Spectroscopy by \citet{williams1977} showed
that the northern component is a strong H$\alpha$ emission line star
while the southern component shows a pure absorption spectrum, and
they suggested spectral types of B4V and B7V, respectively.    
In the near infrared the northern
component has excess emission due to circumstellar dust 
\citep{williams1977}. Optical and/or
NIR photometry is given by \citet{williams1977},
\citet{MarracoForte1978}, 
\citet{chellietal1995} and  2MASS. The \JHKLM \  photometry in
\citet{williams1977} includes both components.
\jhks \  photometry in 2MASS is available only for h4636s 
and only in \J \ for h4636n (the stars are bright and
separated only by 3\farcs7).  
\citet{chellietal1995} were able to provide the \jhks \ magnitude
  differences between the binary components from one dimensional
  specklegraph scan measurement. However, the magnitudes and
magnitude differences from
different dates do not agree. A plausible explanation for this is
that one or both of the binary components are variable. Optical
variability is suspected also by \citet{williams1977}. The 2MASS \jhks
\ magnitudes for h4636s were taken simultaneously and its colour indices
can be calculated. Its position in Fig. \ref{figure:jhk} indicates
strong NIR excess.  Both the components of h4636 are strong X-ray
detections (I\_56 and I\_58, \citetalias{getmanetal2008}).

Star \five \ (\object{J 135742.28-395846.7}), which coincides with the
X-ray source I\_54 \citepalias{getmanetal2008}, lies west of
h4636. The star has \jhks \ colours indicative of NIR excess and a
reddening corresponding to an Av of more than 10 magnitudes if
originating from the CTTS locus.  \citetalias{getmanetal2008} estimate
an Av of 25 magnitudes.  It was the only source with significant
variability during the Chandra observations
\citepalias{getmanetal2008}.  The star is visible in all but the 8
$\mum$ Spitzer images. At 8 $\mum$ the nebulosity at the position of
the star becomes strong and this may be the reason why the star is not
detected. Star \five \ is a Class I candidate.

A faint star, \nnn \ (\object{J 135744.32-395853.5}), is detected 
south of h4636.  The star is within the strong gradient of the bright
h4636 nebulosity and the \J \ magnitude of the source is only an
estimate.  The star is detected in an L band but not in an M band ISAAC
image (Yun 2008, private communication). It is not seen in the Spitzer
images.  The star is a potential member of the CG~12 stellar cluster.

\subsubsection{ \citet{williams1977}  stars 2, 5, 6, and  8 } \label{sec:stars2568} 

\citet{williams1977} stars W77-2 (\object{2MASS J13575601-4004192},
\object{CD--39$^\circ$8583}) and W77-8 (\object{2MASS
  J13572609-3956091}) are not within the NTT SOFI image but their \J,
\H, and \Ks \ magnitudes are available in 2MASS. Star W77-2 is a late
B star \citep{maheswar1996} and W77-8 an A0 type star
\citep{gahm1980}. Both stars are surrounded by optical nebulosities
and are members of the CG~12 stellar cluster.  The SOFI photometry of
star W77-5 (\object{2MASS J13574803-3957295} is in agreement with the
2MASS values even though it lies near the saturation limit.  2MASS
photometry is, however, adopted when calculating the colour indices.
The star is positioned on the late main sequence in
Fig. \ref{figure:jhk}.  This is in line with its spectral type K0
given in \citet{gahm1980} and its location in the UBV colour-colour
diagram  \citep{williams1977}. The star is a foreground object.

Star W77-6 is a binary (\wsixa, \object{J 135744.43-395650.5} and \wsixb,
\object{J 135744.70-395652.8}) with a separation of $\sim$4\arcsec. The
only good quality 2MASS magnitudes are for the brighter binary
component in \H \ and \Ks.  The SOFI (\J-\H) and (\H-\Ks) colour
indices are 0\fm37 and 0\fm25 for the brighter, northern (a) component
(\wsixa, W77-6a) and 1\fm20 and 0\fm75 magnitudes for the southern (b)
component (\wsixb, W77-6b). Because of the brightness of star \wsixa \ in
all three bands and \wsixb \ in the \K \ band they fall on the
nonlinear part of the detector so the magnitudes in these bands and the
colour indices derived are somewhat uncertain. Star W77-6b
is strongly variable, and the photometry above is for the 2005
observations when single night \jhks \ photometry is available. The
\J \ magnitude of star W77-6b varied from 13\fm90 to 14\fm20 between 2005
and the first observation in 2007.  In a single night in 2007 the \J
\ magnitude changed from 14\fm20 to 14\fm84.  This is far more than
can be explained by the uncertainty introduced by the background
nebulosity and detector nonlinearity. The \H \ and \Ks \ magnitudes in
2005 and 2007 are the same within the adopted conservative photometric
uncertainty (0\fm1).  Despite the photometric uncertainties,
the colour indices of star W77-6a and the 2005 star W77-6b are  suited for
the following analysis. The position of star W77-6a adjacent to W77-8 in the
\CCD \ indicates that it is a slightly reddened early A-type star. W77-6b
has a NIR excess which indicates a circumstellar disk. In the
optical the W77-6 binary is associated with a bright nebulous patch. In
the Spitzer images W77-6 displays a complex shape where a nebulosity
varying in shape and size is seen at the position of star
W77-6a. Neither of the W77-6 components was detected in X-rays
\citepalias{getmanetal2008}. W77-6 is further discussed in
Sect. \ref{sec:compare} below.

\subsubsection{ Star 2MASS \object{J13574059-4000590}} \label{sec:starJ13574059-4000590}

 This bright star (2MASS \J, \H, and \Ks \ magnitudes 9.44, 8.24, and
 7.75 magnitudes, respectively) lies near the CG~12~SW core. It is
 outside the analysed SOFI image but is seen at the southern corner in
 Fig. \ref{figure:features}. The (\J-\H) and (\H-\Ks) colours are
 1\fm2 and 0\fm49, and if it were assumed to be  a main sequence star its visual
 extinction Av would be 4\umag \ or more. As the strong reddening
 starts first at the distance of CG~12 \citep{maheswar1996}, the star
 is either embedded in CG~12 or located behind the cloud.  If embedded
 its calculated dereddened absolute \Ks \ magnitude would be
 approximately -2\umag.  The star is either a very early type object
 or, more likely, a late-type giant 
behind the cloud. But spectroscopic
 observations are needed to clarify whether the star is a
 late-type giant or an early type star.  In the Spitzer images the star is
 saturated in all but the 8 $\mum$ band.

\subsubsection{Chandra X-ray detections} \label{chandra} 

\citetalias{getmanetal2008} argue that half of the 128 Chandra X-ray
detections in the direction of CG~12 are likely to be T-Tau stars
associated with CG~12.  The detections within the SOFI image coincide
with stars \two, \three/\four, \five, h4636n/h4636s, \fourone
\ (\object{J 135732.16-395836.3}, I\_36), \eightone \ (\object{J
  135738.21-395947.7}, I\_42) \ and \twooei \ (\object{J
  1357537.4-395730.8}, I\_68). The first six stars have been discussed
above. Of the last three only star \eightone \ has significant
extinction (Av\,$\sim$7\umag) and the other two lie near the main
sequence. The observed apparent magnitudes of these stars are in
agreement with them being at the assumed distance of CG~12 (550 to 630
pc).

Of the remaining five \citetalias{getmanetal2008} X-ray detections in
the SOFI field one (I\_37) coincides with a galaxy and four (I\_31,
I\_40, I\_63 and I\_71) have no NIR counterpart in the SOFI images.
These four are extragalactic X-ray candidates. The
\citetalias{getmanetal2008} X-ray detections I\_23 (\object{2MASS
  J13572321-3957498}), I\_27 (\object{2MASS J13572583-3958372}), I\_29
(\object{2MASSJ13572601-3956271}), I\_47 \ (\object{2MASS
  J13573973-3954035}), and I\_61 (\object{2MASS J13574510-4001071})
coincide with stellar sources at the cropped-off edges of the SOFI
images and are identified in Fig. \ref{figure:features}.  Of these
five stars SOFI three-colour data are available only for I\_29 and
I\_47. Even though the star I\_47 is at the edge of the SOFI images
where reliable photometry is difficult, the photometry is in agreement
with the 2MASS magnitudes (maximum deviation 0\fm07). The situation is
different for I\_29. The SOFI and 2MASS \jhks \ magnitudes are
14\fm10, 13\fm02, 12\fm48 and 14\fm84, 13\fm44, 12\fm63
respectively. The difference is much higher than the one observed for
I\_29, even though the two stars are of similar brightness.  The
observed Chandra source median energy of 1.2 keV, which indicates a
low absorbing column density, was considered unusual by
\citetalias{getmanetal2008} because the 2MASS photometry
((\J-\H)=1\fm44, (\H-\Ks)=0\fm81) infers a visual extinction of $\sim$6\umag,
which is higher than expected as well as  NIR excess. The SOFI photometry
((\J-\H)=1\fm08, (\H-\Ks)=0\fm54) does not indicate NIR excess and the
inferred extinction is lower. The star is probably variable.

 \subsubsection{Potential cluster member stars } \label{potentialmembers} 

 As discussed above, all stars in the SOFI \Ks \ image with apparent or
 de-reddened magnitude brighter than $\sim$15\umag \ are possible
 members of the cluster associated with CG~12. However, for most of
 the stars \jhks \ photometry alone is not sufficient to discriminate
 between cluster members and field stars. NIR excess and an associated
 nebulosity or X-ray activity of some of the stars confirms their
 membership.

The location of many of the stars which are brighter 
than 15\fm0 in \Ks \ (marked with red circles) in the \CCD \ 
(Fig. \ref{figure:jhk}) indicates extinction 
up to a few magnitudes in the
visual, showing that they are either embedded in CG~12 or are
background stars. Some of these could be T-Tauri stars which were not
active during the \citetalias{getmanetal2008} Chandra
observations. The \citetalias{getmanetal2008} objects I\_36 and I\_68
(stars \fourone \ and \twooei) lie near the unreddened main sequence.
A group of bright stars lies near (\J-\H),(\H-\Ks) 1.7,0.8 in the \CCD.
Apart from the X-ray sources I\_41 and I\_42 (stars \two \ and \eightone)
and star \six \ three other stars (76, 93 and 99) belong to the
group. These stars are potential cluster members, but further,
e.g. spectroscopic, observations are needed to confirm the membership.

In addition to stars already discussed above, star 80 is the only star
for which \jhks \ colours indicate clear NIR excess. The star is,
however, much fainter than e.g. the two other stars (\four\  and
\five) with strong NIR excess.

\subsubsection{Visual extinction } \label{sec:extinction}

The NICE method presented in
 \citet{ladalada1994} and the SOFI NIR photometry can be used to
 evaluate the visual extinction within the CG~12 cloud.  However, the
 location of CG~12 high above the Galactic plane is not favourable for
 applying the NICE method. According to the \citet{mattila1980} model
 the expected number of stars behind CG~12 up to a distance of 1 kpc
 is only 19. Like the foreground stars these are of the spectral type G5
 or later.  If the stars are distributed randomly in the SOFI image,
 not many are expected in the direction of the likely positions of the
 highest extinction, \cgs and \cgnp.  Because of the high surface
 brightness in these positions,  the \J \ limiting magnitude is also
 lower than the value 21\fm0 quoted for the SOFI image as a whole in
 Sect. \ref{sec:observations}.  One would therefore not expect \J
 \ band detections of background stars lying farther than 1 kpc in the
 directions of high ($>$20\umag) extinction.  After eliminating the
 known CG~12 cluster members and the few assumed foreground stars, the
 visual extinction in the cloud could be estimated.  However, as
 argued above, the SOFI data are not deep enough to probe the visual
 extinction through the two densest cloud cores, \cgn and \cgsp.  The
 observed extinction reaches to $\sim$20\umag\ around \cgn \ and 
 $\sim$15\umag \ around \cgsp. The lower measured extinction towards
 \cgs is most probably due to the very bright surface emission around
 h4636 and the cone (Fig. \ref{figure:CG12_3col}). which is the reason 
for the limiting magnitude to be even lower than in the direction of \cgnp.
 There are no background stars bright enough to be observed through
 the core centres, and therefore only a lower limit for the visual
 extinction is obtained. As predicted, the stars suffering the highest
 extinction are \H,\Ks \ detections only. The \CCD \ in
 Fig. \ref{figure:jhk} indicates that the extinction does not reach
 zero within the SOFI image. This is to be expected as the observed
 optical CG~12 surface brightness extends far outside the SOFI image
 (see, e.g., Fig. 8 in Paper 1).

\subsection{Surface brightness } \label{sec:surface}

 Apart from the two nebulous bright binaries in the SOFI field the most
 intriguing feature in the image is the cone-shaped nebulosity SW from
 h4636. Indication of the bright cone opening to the North is already
 faintly seen in the 2MASS images, but the fainter counter cone opening
 to the South is only seen in the SOFI \Ks \ image (see
 Fig. \ref{figure:KsHIRES}). Together the cones form an
 hourglass-shaped object with a bright point at the ``waist''. This
 object is listed as a 2MASS source \object{2MASS J13574147-3959252}. This
 source is not stellar-like in the SOFI \jhks \ images but blends into
 the surface brightness.  The cone and the central source are visible
 in the Spitzer 3.6 and 4.5 $\mum$ images.  The central source is
 still visible in the 6 $\mum$ image but the cone blends into the
 background. At 8 $\mum$ only background nebulosity is seen.  The
 cone is projected on the \dcopl core and the 1.2\,mm continuum source
 \citepalias{haikalaolberg2007, haikalaetal2006} and it coincides with
 the centre of the collimated molecular outflow detected by
 \citet{white1993}.  The hourglass is therefore most probably a cavity
 cleared by the outflow where the walls are reflecting light from the
 outflow central source.  A long
 cavity opening north of the bright cone is best seen in the \J
 \ image. It corresponds to a cavity vacated by the outflow into the
 less tenuous part of the dust/molecular cloud. A similar but less
 pronounced feature is seen also south of the hourglass.

  The north-south dark lane between star \two \ and the binary 
\three/\four \ is likely to be a shadow cast by the star \four
  \ circumstellar disk and not due to obscuration in the dust
  cloud. Such shadows were first recognised by \citet{hodappetal2004}
  in NGC1333. The position of the shadow with respect to the probable
  circumstellar disk around star \four \ corresponds to the
  theoretical \citet{pontoppidanetal2005} scenario 2a in which a
  shadow cast by the disk is seen on the surface of an adjacent dust
  cloud.

 The NIR surface brightness is a powerful tool for studying the
 structure of isolated dark cloud cores illuminated by the diffuse
 interstellar radiation field (ISRF). For an ideal spherical dust
 cloud the \J, \H,  and \K \ surface brightness due to reflected ISRF
 forms concentric rings where the radius gets smaller  the longer the
 wavelength. The inner radius of each ring corresponds to the optical
 depth of approximately 1. The phenomenon has been observed  e.g.  by
 \citet{lehtinenmattila1996} and discussed in detail in
 \citet{juvelaetal2008}. CG~12 does not correspond to this ideal case
 as because apart from the ISRF the cloud also shows a strong internal source of
 light (h4636). 
The binary in \cgn (stars \three \ and \four) and
 W77-6 also act as smaller and more localised sources of light. Offset from
 these light sources there are three localised positions where the
 long wavelength emission becomes strong, and the surface brightness
 turns redder in Fig. \ref{figure:CG12_3col}. These positions are
 south of the cone, east of the binary \three/\four \ and around star
 \one. The first two coincide with the millimetre continuum sources
 \citepalias{haikalaolberg2007}, exactly in positions where one would
 expect the phenomenon.

\subsubsection{Spitzer NIR imaging } \label{sec:spitzer}

The Spitzer imaging covers a much larger area than the SOFI
observations presented in this paper. In the future the longer wavelength IRAC
observations will 
provide additional information for
stars inside the SOFI field and, will in particular allow the
construction of a meaningful NIR spectral energy distribution from 1
$\mum$ \ to 8 $\mum$ \ for many stars. This will e.g. permit the
separation of  the background objects from the list of potential member
stars in Table \ref{table:jhkphotometry} as well as a separation 
of the Class~I
and II objects. Apart from the objects already discussed above, the new
Spitzer imaging does not provide new obvious bright CG~12 cluster
member candidates in the SOFI field.

The probable driving source of the collimated molecular outflow
\citep{white1993} is better resolved in the Spitzer data than in the
SOFI \jhks \ images, where the object is heavily blended with the
background nebulosity.  A major part of the object in the waist of the
``hourglass'' in the SOFI \H \ and \Ks \ images is not direct emission
from the central object, but light reflected from its surroundings.

In the SOFI data the binary W77-6 does have a surrounding nebulosity, but
the binary itself can be well resolved. The Spitzer data show a
complicated nebulous structure which varies with wavelength and a
bright stellar-like source north of W77-6b. The source is very bright in
the 5.8\,$\mum$ \ image.

\subsection{IRAS point sources} \label{sec:compare}  

CG~12 was observed by the IRAS satellite at wavelengths between 12 and 100
$\mum$.  The spatial resolution of the original IRAS images can be
enhanced using HIRES processing which uses the maximum correlation
technique \citep{aumanetal1990}.  The early HIRES processed
maps had often wrong coordinate zero points. The HIRES maps
presented in \citet{white1993} are offset to the South from the
correct position.  The spatial resolution of the 60 and 100 $\mum$
IRAS maps, even if enhanced with the HIRES processing, is not sufficient
to resolve details in CG~12 \citep[see][Fig.~3]{white1993}.
The mid- and far-IR emission in CG~12 is dominated by the strong point
source \object{IRAS 13547--3944} near h4636. The second much fainter
point source in CG 12 is \object{IRAS 13546--3941}.  Its nominal
position is slightly offset to the southeast from the \ceo maximum
\cgn and a millimetre continuum source \citepalias{haikalaetal2006}.
The IRAS small extended source \object{X1354-397} is seen in the
direction of the \dcopl core.

Two further IRAS point sources are associated with CG~12.  \object{IRAS
13543--3941} lies west of \cgn towards the  star W77-8
and the optical nebulosity \object{Bernes 146}. \object{IRAS 13549--3950}  
south of the 
SOFI image lies towards the star W77-2 and an optical  reflection
nebulosity.

\subsubsection{IRAS 13547--3944} \label{sec:IRAS13547--3944} 

\begin{figure} \centering \includegraphics 
[width=8cm, angle=-90.0]{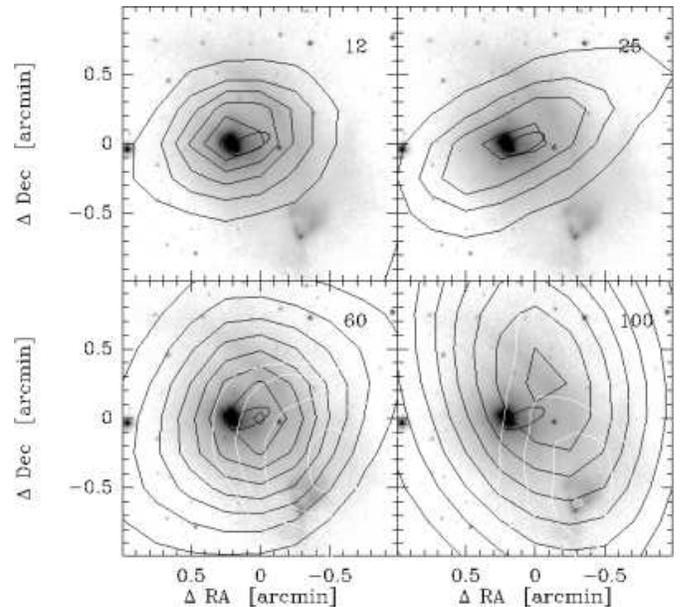}
\caption{Contour maps (arbitrary contour values) of the HIRES
  processed IRAS maps superposed on the SOFI \Ks \ band image. The
  IRAS 13547--3944 positional uncertainty ellipse is also shown. The map
  wavelength is shown in microns in the upper right corner of each
  panel. Superposed using white contours on the 60 $\mum$ panel are the
  \ceo (2-1) integrated emission contours
  \citepalias{haikalaolberg2007} and on the 100 $\mum$ panel the 1.2 mm
    continuum emission (Haikala 2010, in preparation).  The offset in
    arcminutes from the IRAS 13547--3944 point source position is shown
    on the axis.}
 \label{figure:KsHIRES}
\end{figure}

IRAS 13547--3944 has been frequently associated with the binary
h4636. However, the binary lies east of the point source outside its
positional uncertainty ellipse.  The observed emission line spectrum
of h4636n and the infrared excess deduced from the \JHKL \ photometry
containing both the n and s components led \citet{williams1977} to
model the binary with a B7 star (h4636s) plus a B4 star (h4636n) with
a warm (1600K) circumstellar shell.  The point source fluxes are 7.8,
8.9, 67.5, 201.9 Jy at 12, 25, 60 and 100 $\mum$, respectively.  The
IRAS fluxes at 12 and 25 $\mum$ could be appropriate for a warm
circumstellar shell. In the ISO-LWS FIR colour-colour diagram, IRAS
13547--3944 lies in the region occupied by Class~0 sources well offset
from HAEBE objects \citep{pezzutto2002}.  This and the IRAS~60 and
100\,$\mum$ fluxes point at a cold cloud core and not to a warm
circumstellar shell.  The highest contours of the HIRES processed IRAS
emission in the 12 to 100 $\mum$ bands is shown superposed on the SOFI
\Ks \ band image in Fig. \ref{figure:KsHIRES}. The white contours in
the 60 $\mum$ and the 100 $\mum$ panels trace the \ceo (2-1) integrated 
emission
\citepalias{haikalaolberg2007} and the 1.2\,mm continuum emission
(Haikala 2010, in preparation) respectively.  The IRAS
13547--3944 point source position uncertainty ellipse is drawn in all
the panels.

 Fig. \ref{figure:KsHIRES} is in accord with a confused situation
 where two independent sources, one warm and one cold, are separated
 by less than the IRAS beam at 12\,$\mum$. Given the IRAS low spatial
 resolution (30$''$ at 12\,$\mum$ and about 2$'$ at 100\,$\mum$), h4636n
 has been in the IRAS beam and has contributed to the point source
 flux and to also to the point source position at least at
 12\,$\mum$. The hypothetical
cold source would be well within the IRAS
 FIR (and ISO-LWS) beams.  The 1.2\,mm continuum maximum,
 which coincides with the \dcopl core 
and the NIR cone, lies SW of the
 nominal position of the point source and from the maxima of IR
 emission in all the IRAS bands. However, the 1.2\,mm continuum
 emission is elongated towards the IRAS 100\,$\mum$ maximum.  It is not
 likely that star \five \ is the cold IRAS 13547--3944 source. Star
 \five \ could contribute to the 12 and 25\,$\mum$ IRAF fluxes but the
 ISO-LWS data point at a cold source at an earlier phase of
 evolution.

\subsubsection{IRAS  13546--3941} \label{sec:IRAS13546--3941}
\begin{figure} \centering \includegraphics 
[width=6cm, angle=-90.0]{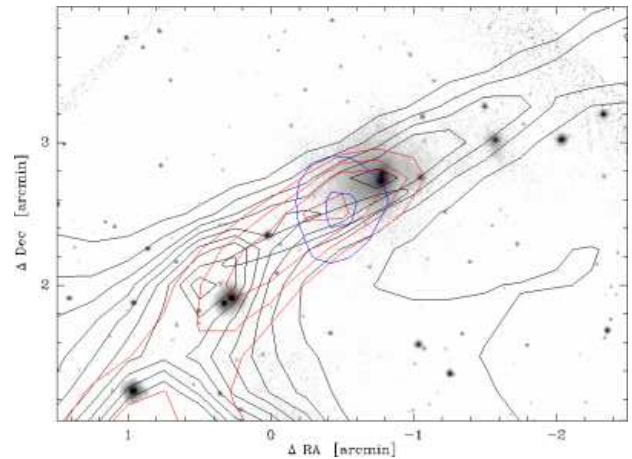}
\caption{Contour maps (arbitrary contour values) of the HIRES
  processed IRAS 12 $\mum$ (black contours) and 25 $\mum$ (red
  contours) maps of the \cgn region superposed on the SOFI \Ks \ band
  image. The blue contour traces the maximum of the 1.2 mm
    continuum emission (Haikala 2010, in preparation).
 The IRAS  13546--3941  positional uncertainty ellipse is indicated by the ellipse.
  Axis offsets are as in Fig. \ref{figure:KsHIRES}}
 \label{figure:KsHIRESnorth}
\end{figure}

This point source, which lies near the tip of the globule and between
the binaries W77-6a/b and 86/87,
 is a 25\,$\mum$-detection only (1.38 Jy) and
  only upper limits are given in the three other IRAS bands.   HIRES
processed IRAS 12 (in red) and 25\,$\mum$ (in black) contour maps of
\cgn superposed on the SOFI \Ks \ image are shown in
Fig. \ref{figure:KsHIRESnorth}. The blue contours trace the 1.2\,mm
continuum emission maximum (Haikala 2010, in preparation).  The
positional uncertainty ellipse of IRAS 13546--3941 is also
shown. The 12\,$\mum$ map has two maxima, one around the binary
\three/\four \ and another just to the side of the binary W77-6. Both
maxima lie outside the IRAS point source positional uncertainty error
ellipse.  In the 25 $\mum$ map the emission is elongated and
encompasses both binaries, but the maximum lies between the binaries
near the \ceo maximum \cgn and the 1.2\,mm continuum source which is
indicated with the blue contour in Fig. \ref{figure:KsHIRESnorth}.
There is no clear indication of 12 or 25 $\mum$
emission in the direction of star \one.
  
 The IR excess of star \four \ in the \CCD \ and the probable disk
 shadow west of it is a strong indication of a circumstellar
 disk. One would expect the disk to be visible at 12\,$\mum$.  The
 \jhks \ colours of W77-6b indicate NIR excess and a circumstellar
 disk or shell. A strong nebulosity surrounds W77-6 in the Spitzer 5.8
 and 8 $\mum$ \ images, possibly indicating a further source which is
 not seen in the NIR SOFI images. This putative source and/or the
 W77-6b circumstellar shell/disk could explain the 12\,$\mum$ HIRES
 maximum near the star. The HIRES 25\,$\mum$\ maximum, which nearly
 coincides with the IRAS 13546--3941 point source position, lies
 between the binaries and coincides with the \ceo maximum \cgnp, and
 the 1.2\,mm continuum core.

 The location of the HIRES 12\,$\mum$\ maxima with respect to the 
NIR excess stars \four \ 
and W77-6b indicates strongly that these are 
real 12\,$\mum$ sources and not artifacts of the HIRES processing.
 Future mid- and far-IR observations with
 better spatial resolution than was possible with the IRAS satellite
 are, however, necessary to confirm this.

\section{Summary and conclusions} \label{sec:summary} 

The head of the cometary globule CG~12 has been observed in the \J, \H,
and \Ks \ bands.  The new NIR photometry combined with 
already existing data at other wavelengths allows the analysis of the star
formation and the content of the associated stellar cluster in more
detail than was possible before.

 Star formation in CG 12 (within the region imaged in \jhks \ with
 SOFI) is concentrated on and near the two strong \ceo maxima \cgs and
 \cgnp. Apart from the known members of the associated stellar cluster
 several new, embedded cluster members were detected as
 well. Because the SOFI images are much deeper than the available
 2MASS survey data it is possible to define more accurately the NIR
 colours of the already known embedded member stars which were near
 the 2MASS limit.

Apart from the stellar photometry, the SOFI imaging allows the study of the
cloud surface brightness due to scattered light. Scattered light
permits us to pinpoint the central source of the highly collimated molecular
outflow detected by \citet{white1993} and to detect the probable
shadow cast by the circumstellar disk around star \four.
  
Using NIR \J,\H,\Ks \ photometry alone is not in all cases sufficient to
decide whether some stars are indeed members of the CG~12 stellar
cluster. In some cases an associated nebulosity can
confirm the membership. For four stars the observed X-ray activity
\citepalias{getmanetal2008} indicates that the stars are likely members.

 The conclusions are the following:

 1. Seven embedded (proto)stars (\one, \two, \three, \four, \five,
 \six\ and W77-6b) are identified.  The visual extinction of these
 stars ranges from 10 to more than 30 magnitudes.  The stars \three
 \ and \four \ were already known members. Stars \two, \ \three,
 \ \four \ and \five \
  are also  detected in X-rays \citepalias{getmanetal2008}.

 2. Stars \one, \four, \five, \six, and W77-6b have infrared excess
 indicating a circumstellar shell or disk. The shadow of the probable
 disk around star \four \ is seen in the cloud surface brightness.

3. Scattered light from the probable central source of the 
collimated outflow detected by
\citet{white1993} as well as an hourglass-shaped cavity in the
parental molecular cloud formed by the outflow are detected. Apart from  the
hourglass  a larger scale cavity cleared by the outflow to the
less tenuous part of the molecular cloud is also seen. The stellar source 
is better resolved in the Spitzer IRAC images at 3.4, 4.5 and 5.8 $\mum$ \ 
than in the SOFI \H \ and \Ks \ images, where it blends into the
bright background nebulosity.

{\bf 4.} The \citet{williams1977} star W77-6 is a binary with a separation of
$\sim$4\arcsec.  The brighter component, W77-6a is
an early A type star with a visual extinction of $\sim$3
magnitudes. The fainter component,  W77-6b, has  NIR excess
indicating  circumstellar material. The W77-6b \J \ magnitude changed by 0\fm6
between two observing nights. The
Spitzer NIR imaging suggests that a third source and an associated
nebulosity not visible in the \jhks \  imaging may be present.

 5. HIRES enhanced IRAS 12 and 25\,$\mum$ \ maps combined with
  IRAS fluxes and the NIR imaging suggest that the IRAS point source
  13547--3944 consists of two separate sources. The 12 and 25 $\mum$
  \ emission originates in  the h4636n component and the longer wavelengths from an
  adjacent, embedded cold cloud source.  The 12 $\mum$ \ HIRES map of
  IRAS 13546--3941 resolves into two previously undetected sources
  coinciding with the binary stars W77-6a/b and \three/\four. The
  HIRES 25 $\mum$ maximum lies between the 12 $\mum$ \ sources near
  the IRAS 13546--3941 nominal position and the \ceo maximum and a mm
  continuum source. The 25\,$\mum$\ detection could be a source
  embedded in the continuum core. 

6. The SOFI data are not deep enough to probe the visual extinction
through the two dense cloud cores in the direction of \cgn and
\cgsp. The observed maximum extinction around \cgn is $\sim$20\umag,
and $\sim$15\umag \ around \cgsp. No background stars are seen
through the core centres.

\begin{acknowledgements}
 Joao Yun is thanked for providing the ISAAC K and L images of \cgsp. It is a
 pleasure to thank the NTT team for the support during the two
 observing runs. BR acknowledges support by the National Aeronautics
 and Space Administration through the NASA Astrobiology Institute under
 Cooperative Agreement No. NNA04CC08A issued through the Office of Space
 Science. This research has made use of the SIMBAD database,
operated at CDS, Strasbourg, France, and of NASA's Astrophysics Data System
Bibliographic Services.
\end{acknowledgements}
\bibliographystyle{aa}
\bibliography{cg12_sofi_v4.bib}
\listofobjects
\Online
\onllongtab{1}{
\begin{longtable}{lcccccccc}     
\caption{\label{table:jhkphotometry} Stellar \jhks \ photometry.} \\
\hline\hline
     Star  &      ID       &  J2000             &\J   &  \H & \Ks & (\J-\H)& (\H-\Ks)&  Comment\\
     \#    &               & hhmmss.ss ddmmss.s & mag & mag & mag &   mag  &    mag&       \\   
\hline   
\endfirsthead
\caption{Continued.} \\
\hline
     Star  &      ID       &  J2000             &\J   &  \H & \Ks & (\J-\H)& (\H-\Ks)&  Comment\\
     \#    &               & hhmmss.ss ddmmss.s & mag & mag & mag &   mag  &    mag  &     \\  
\hline
\endhead
\hline
\endfoot
\hline
\endlastfoot
    1&   J 135725.43-395731.1&13 57 25.43 -39 57 31.1& 14.61$\pm$0.01&  13.54$\pm$0.01&  13.17$\pm$0.01&   1.07&   0.37&          \\
    2&   J 135725.51-395715.7&13 57 25.51 -39 57 15.7& 19.93$\pm$0.04&  18.84$\pm$0.05&  18.41$\pm$0.04&   1.09&   0.43&          \\
    3&   J 135726.07-395715.6&13 57 26.07 -39 57 15.6& 16.63$\pm$0.01&  15.85$\pm$0.01&  15.60$\pm$0.01&   0.78&   0.25&          \\
    4&   J 135726.12-395723.8&13 57 26.12 -39 57 23.8& 19.17$\pm$0.02&  18.46$\pm$0.04&  18.11$\pm$0.04&   0.71&   0.35&          \\
    5&   J 135726.41-395647.8&13 57 26.41 -39 56 47.8& 15.35$\pm$0.01&  14.36$\pm$0.01&  14.03$\pm$0.01&   0.99&   0.33&          \\
    6&   J 135726.84-395800.4&13 57 26.84 -39 58 00.4& 19.56$\pm$0.04&  18.68$\pm$0.05&  18.35$\pm$0.04&   0.88&   0.33&          \\
    7&   J 135727.03-395759.2&13 57 27.03 -39 57 59.2& 19.07$\pm$0.02&  18.46$\pm$0.03&  18.25$\pm$0.04&   0.61&   0.21&          \\
    8&   J 135727.36-395819.1&13 57 27.36 -39 58 19.1& 19.29$\pm$0.02&  18.62$\pm$0.04&  18.18$\pm$0.04&   0.67&   0.44&          \\
    9&   J 135727.42-395717.8&13 57 27.42 -39 57 17.8& 17.84$\pm$0.01&  16.93$\pm$0.01&  16.61$\pm$0.01&   0.91&   0.32&          \\
   10&   J 135727.54-395722.7&13 57 27.54 -39 57 22.7& 16.92$\pm$0.01&  16.35$\pm$0.01&  16.04$\pm$0.01&   0.57&   0.31&          \\
   11&   J 135727.55-395803.2&13 57 27.55 -39 58 03.2& 19.96$\pm$0.04&  19.33$\pm$0.06&  18.83$\pm$0.06&   0.63&   0.50&          \\
   12&   J 135727.66-395809.8&13 57 27.66 -39 58 09.8& 19.93$\pm$0.04&  19.24$\pm$0.06&  18.66$\pm$0.07&   0.69&   0.58&          \\
   13&   J 135728.25-395739.3&13 57 28.25 -39 57 39.3& 19.13$\pm$0.02&  18.04$\pm$0.03&  17.73$\pm$0.03&   1.09&   0.31&          \\
   14&   J 135728.29-395745.9&13 57 28.29 -39 57 45.9& 17.73$\pm$0.01&  17.09$\pm$0.01&  16.92$\pm$0.01&   0.64&   0.17&          \\
   15&   J 135728.51-395736.9&13 57 28.51 -39 57 36.9& 20.39$\pm$0.07&  19.62$\pm$0.08&  19.23$\pm$0.08&   0.77&   0.39&          \\
   16&   J 135728.75-395729.6&13 57 28.75 -39 57 29.6& 19.64$\pm$0.03&  18.89$\pm$0.05&  18.87$\pm$0.07&   0.75&   0.02&          \\
   17&   J 135728.80-395815.0&13 57 28.80 -39 58 15.0& 19.35$\pm$0.02&  18.67$\pm$0.04&  18.29$\pm$0.04&   0.68&   0.38&          \\
   18&   J 135728.97-395710.8&13 57 28.97 -39 57 10.8& 19.16$\pm$0.02&  18.29$\pm$0.03&  17.96$\pm$0.03&   0.87&   0.33&          \\
   19&   J 135729.10-395624.5&13 57 29.10 -39 56 24.5& 20.42$\pm$0.06&  19.09$\pm$0.06&  18.56$\pm$0.05&   1.33&   0.53&          \\
   20&   J 135729.20-395832.7&13 57 29.20 -39 58 32.7& 19.26$\pm$0.03&  18.30$\pm$0.04&  17.99$\pm$0.03&   0.96&   0.31&          \\
   21&   J 135729.23-395750.8&13 57 29.23 -39 57 50.8& 11.85$\pm$0.01&  11.05$\pm$0.01&  10.85$\pm$0.01&   0.80&   0.20&          \\
   22&   J 135729.81-395711.3&13 57 29.81 -39 57 11.3& 18.29$\pm$0.01&  17.33$\pm$0.01&  16.99$\pm$0.01&   0.96&   0.34&          \\
   23&   J 135729.94-395758.3&13 57 29.94 -39 57 58.3& 19.48$\pm$0.02&  18.86$\pm$0.05&  18.60$\pm$0.06&   0.62&   0.26&          \\
   24&   J 135729.97-395808.4&13 57 29.97 -39 58 08.4& 18.93$\pm$0.02&  18.26$\pm$0.03&  17.99$\pm$0.03&   0.67&   0.27&          \\
   25&   J 135730.12-395759.9&13 57 30.12 -39 57 59.9& 19.64$\pm$0.04&  18.83$\pm$0.06&  18.53$\pm$0.06&   0.81&   0.30&          \\
   26&   J 135730.26-395631.6&13 57 30.26 -39 56 31.6& 20.00$\pm$0.04&  18.68$\pm$0.04&  18.00$\pm$0.03&   1.32&   0.68&          \\
   27&   J 135730.33-395649.1&13 57 30.33 -39 56 49.1& 19.57$\pm$0.04&  18.52$\pm$0.04&  18.16$\pm$0.04&   1.05&   0.36&          \\
   28&   J 135730.51-395645.9&13 57 30.51 -39 56 45.9& 19.66$\pm$0.03&  18.57$\pm$0.04&  18.09$\pm$0.03&   1.09&   0.48&          \\
   29&   J 135730.53-395659.6&13 57 30.53 -39 56 59.6& 19.42$\pm$0.03&  18.31$\pm$0.03&  17.79$\pm$0.03&   1.11&   0.52&          \\
   30&   J 135730.65-395704.4&13 57 30.65 -39 57 04.4& 15.98$\pm$0.01&  14.85$\pm$0.01&  14.50$\pm$0.01&   1.13&   0.35&          \\
   31&   J 135730.71-395649.2&13 57 30.71 -39 56 49.2& 20.50$\pm$0.10&  19.20$\pm$0.09&  18.63$\pm$0.06&   1.30&   0.57&          \\
   32&   J 135730.74-395707.4&13 57 30.74 -39 57 07.4& 18.81$\pm$0.02&  17.84$\pm$0.02&  17.51$\pm$0.02&   0.97&   0.33&          \\
   33&   J 135730.79-395803.7&13 57 30.79 -39 58 03.7& 19.87$\pm$0.03&  18.91$\pm$0.05&  18.54$\pm$0.06&   0.96&   0.37&          \\
   34&   J 135730.90-395742.6&13 57 30.90 -39 57 42.6& 20.64$\pm$0.07&  19.85$\pm$0.10&  19.20$\pm$0.10&   0.79&   0.65&          \\
   35&   J 135731.02-395836.0&13 57 31.02 -39 58 36.0& 19.48$\pm$0.03&  18.50$\pm$0.04&  18.27$\pm$0.04&   0.98&   0.23&          \\
   36&   J 135731.04-395755.8&13 57 31.04 -39 57 55.8& 15.63$\pm$0.01&  14.87$\pm$0.01&  14.66$\pm$0.01&   0.76&   0.21&          \\
   37&   J 135731.06-395636.3&13 57 31.06 -39 56 36.3& 19.95$\pm$0.04&  18.39$\pm$0.04&  17.82$\pm$0.03&   1.56&   0.57&          \\
   38&   J 135731.46-395734.6&13 57 31.46 -39 57 34.6& 19.27$\pm$0.02&  18.31$\pm$0.03&  17.90$\pm$0.03&   0.96&   0.41&          \\
   39&   J 135731.93-395812.9&13 57 31.93 -39 58 12.9& 19.73$\pm$0.04&  19.08$\pm$0.06&  18.88$\pm$0.06&   0.65&   0.20&          \\
   40&   J 135732.09-395804.2&13 57 32.09 -39 58 04.2& 18.90$\pm$0.02&  18.08$\pm$0.02&  17.89$\pm$0.03&   0.82&   0.19&          \\
   41&   J 135732.16-395836.3&13 57 32.16 -39 58 36.3& 14.77$\pm$0.01&  14.21$\pm$0.01&  14.03$\pm$0.01&   0.56&   0.18&1, I\_36  \\
   42&   J 135732.23-395806.3&13 57 32.23 -39 58 06.3& 20.34$\pm$0.04&  19.64$\pm$0.09&  19.21$\pm$0.10&   0.70&   0.43&          \\
   43&   J 135732.32-395543.6&13 57 32.32 -39 55 43.6& 13.89$\pm$0.01&  13.03$\pm$0.01&  12.75$\pm$0.01&   0.86&   0.28&          \\
   44&   J 135732.81-395806.8&13 57 32.81 -39 58 06.8& 17.20$\pm$0.01&  16.50$\pm$0.01&  16.26$\pm$0.01&   0.70&   0.24&          \\
   45&   J 135732.82-395745.8&13 57 32.82 -39 57 45.8& 20.75$\pm$0.05&  20.09$\pm$0.10&  19.44$\pm$0.10&   0.66&   0.65&          \\
   46&   J 135733.08-395905.1&13 57 33.08 -39 59 05.1& 20.18$\pm$0.05&  18.82$\pm$0.04&  18.44$\pm$0.04&   1.36&   0.38&          \\
   47&   J 135733.21-395848.1&13 57 33.21 -39 58 48.1& 17.43$\pm$0.01&  16.79$\pm$0.01&  16.35$\pm$0.01&   0.64&   0.44&          \\
   48&   J 135733.34-395803.4&13 57 33.34 -39 58 03.4& 20.31$\pm$0.05&  19.47$\pm$0.10&  19.29$\pm$0.09&   0.84&   0.18&          \\
   49&   J 135733.34-395743.3&13 57 33.34 -39 57 43.3& 16.09$\pm$0.01&  15.11$\pm$0.01&  14.77$\pm$0.01&   0.98&   0.34&          \\
   50&   J 135733.56-395827.0&13 57 33.56 -39 58 27.0& 20.21$\pm$0.05&  19.12$\pm$0.05&  18.67$\pm$0.05&   1.09&   0.45&          \\
   51&   J 135733.58-395710.5&13 57 33.58 -39 57 10.5& 19.25$\pm$0.02&  17.97$\pm$0.02&  17.29$\pm$0.02&   1.28&   0.68&          \\
   52&   J 135733.71-395537.2&13 57 33.71 -39 55 37.2& 21.09$\pm$0.11&  18.83$\pm$0.05&  18.00$\pm$0.03&   2.26&   0.83&          \\
   53&   J 135733.95-395627.9&13 57 33.95 -39 56 27.9& 19.71$\pm$0.03&  17.84$\pm$0.02&  17.06$\pm$0.01&   1.87&   0.78&          \\
   54&   J 135734.11-395914.1&13 57 34.11 -39 59 14.1& 19.19$\pm$0.02&  17.79$\pm$0.02&  17.25$\pm$0.02&   1.40&   0.54&          \\
   55&   J 135734.14-395850.2&13 57 34.14 -39 58 50.2& 19.80$\pm$0.04&  18.78$\pm$0.05&  18.47$\pm$0.04&   1.02&   0.31&          \\
   56&   J 135734.44-395646.5&13 57 34.44 -39 56 46.5& 20.92$\pm$0.14&  19.34$\pm$0.10&  18.63$\pm$0.07&   1.58&   0.71&          \\
   57&   J 135734.64-395828.3&13 57 34.64 -39 58 28.3& 18.55$\pm$0.01&  17.54$\pm$0.02&  17.15$\pm$0.02&   1.01&   0.39&          \\
   58&   J 135734.76-395543.8&13 57 34.76 -39 55 43.8& 21.62$\pm$0.32&  16.65$\pm$0.01&  13.89$\pm$0.01&   4.97&   2.76& 1        \\
   59&   J 135734.88-395620.1&13 57 34.88 -39 56 20.1& 22.19$\pm$0.24&  19.72$\pm$0.08&  18.50$\pm$0.05&   2.47&   1.22&          \\
   60&   J 135734.92-395827.1&13 57 34.92 -39 58 27.1& 20.10$\pm$0.04&  18.95$\pm$0.05&  18.52$\pm$0.05&   1.15&   0.43&          \\
   61&   J 135735.14-395529.7&13 57 35.14 -39 55 29.7& 16.51$\pm$0.01&  15.93$\pm$0.01&  15.66$\pm$0.01&   0.58&   0.27&          \\
   62&   J 135735.56-395902.1&13 57 35.56 -39 59 02.1& 18.80$\pm$0.02&  17.85$\pm$0.02&  17.44$\pm$0.02&   0.95&   0.41&          \\
   63&   J 135735.65-395855.0&13 57 35.65 -39 58 55.0& 21.32$\pm$0.10&  20.08$\pm$0.10&  19.86$\pm$0.13&   1.24&   0.22&          \\
   64&   J 135735.69-395635.4&13 57 35.69 -39 56 35.4& 20.07$\pm$0.05&  17.98$\pm$0.02&  17.04$\pm$0.01&   2.09&   0.94&          \\
   65&   J 135735.97-395705.7&13 57 35.97 -39 57 05.7& 20.15$\pm$0.05&  18.32$\pm$0.03&  17.48$\pm$0.02&   1.83&   0.84&          \\
   66&   J 135736.29-395800.4&13 57 36.29 -39 58 00.4& 20.45$\pm$0.05&  19.25$\pm$0.06&  18.60$\pm$0.06&   1.20&   0.65&          \\
   67&   J 135736.42-395722.3&13 57 36.42 -39 57 22.3& 14.93$\pm$0.01&  14.31$\pm$0.01&  14.12$\pm$0.01&   0.62&   0.19&          \\
   68&   J 135736.58-395712.1&13 57 36.58 -39 57 12.1& 21.07$\pm$0.09&  19.46$\pm$0.09&  18.70$\pm$0.07&   1.61&   0.76&          \\
   69&   J 135737.01-395519.1&13 57 37.01 -39 55 19.1& 22.02$\pm$0.33&  19.01$\pm$0.05&  17.85$\pm$0.03&   3.01&   1.16&          \\
   70&   J 135737.18-395735.4&13 57 37.18 -39 57 35.4& 20.22$\pm$0.06&  19.18$\pm$0.07&  18.67$\pm$0.06&   1.04&   0.51&          \\
   71&   J 135737.37-395711.9&13 57 37.37 -39 57 11.9& 20.27$\pm$0.06&  18.52$\pm$0.04&  17.70$\pm$0.03&   1.75&   0.82&          \\
   72&   J 135737.43-395534.1&13 57 37.43 -39 55 34.1& 20.39$\pm$0.07&  17.59$\pm$0.02&  16.41$\pm$0.01&   2.80&   1.18&          \\
   73&   J 135737.47-395928.3&13 57 37.47 -39 59 28.3& 21.67$\pm$0.15&  19.82$\pm$0.12&  19.02$\pm$0.09&   1.85&   0.80&          \\
   74&   J 135737.49-395559.5&13 57 37.49 -39 55 59.5& 17.62$\pm$0.01&  15.77$\pm$0.01&  14.93$\pm$0.01&   1.85&   0.84& 1, I\_41 \\
   75&   J 135737.52-395447.5&13 57 37.52 -39 54 47.5& 21.10$\pm$0.10&  19.61$\pm$0.12&  18.98$\pm$0.07&   1.49&   0.63&          \\
   76&   J 135737.57-395710.2&13 57 37.57 -39 57 10.2& 16.56$\pm$0.01&  14.95$\pm$0.01&  14.26$\pm$0.01&   1.61&   0.69& 2        \\
   77&   J 135737.71-400008.7&13 57 37.71 -40 00 08.7& 20.67$\pm$0.07&  19.61$\pm$0.09&  18.94$\pm$0.06&   1.06&   0.67&          \\
   78&   J 135737.71-395718.4&13 57 37.71 -39 57 18.4& 21.03$\pm$0.12&  19.34$\pm$0.09&  18.68$\pm$0.06&   1.69&   0.66&          \\
   79&   J 135737.78-395738.0&13 57 37.78 -39 57 38.0& 19.68$\pm$0.03&  18.38$\pm$0.04&  18.06$\pm$0.04&   1.30&   0.32&          \\
   80&   J 135737.88-395508.0&13 57 37.88 -39 55 08.0& 20.60$\pm$0.08&  18.65$\pm$0.04&  17.32$\pm$0.02&   1.95&   1.33&          \\
   81&   J 135737.98-395759.0&13 57 37.98 -39 57 59.0& 15.36$\pm$0.01&  14.75$\pm$0.01&  14.52$\pm$0.01&   0.61&   0.23&          \\
   82&   J 135738.07-395835.9&13 57 38.07 -39 58 35.9& 20.54$\pm$0.06&  19.11$\pm$0.05&  18.57$\pm$0.05&   1.43&   0.54&          \\
   83&   J 135738.21-395947.7&13 57 38.21 -39 59 47.7& 15.96$\pm$0.01&  14.29$\pm$0.01&  13.62$\pm$0.01&   1.67&   0.67& 1, I\_42 \\
   84&   J 135738.30-395846.0&13 57 38.30 -39 58 46.0& 21.35$\pm$0.11&  19.89$\pm$0.09&  19.15$\pm$0.07&   1.46&   0.74&          \\
   85&   J 135738.75-400015.8&13 57 38.75 -40 00 15.8& 20.60$\pm$0.08&  19.24$\pm$0.06&  18.46$\pm$0.05&   1.36&   0.78&          \\
   86&   J 135738.92-395558.7&13 57 38.92 -39 55 58.7& 15.25$\pm$0.01&  12.65$\pm$0.01&  11.38$\pm$0.01&   2.60&   1.27& 1        \\
   87&   J 135738.95-395600.7&13 57 38.95 -39 56 00.7& 16.35$\pm$0.01&  12.88$\pm$0.01&  10.84$\pm$0.01&   3.47&   2.04& 1        \\
   88&   J 135738.98-395505.3&13 57 38.98 -39 55 05.3& 21.04$\pm$0.09&  19.29$\pm$0.06&  18.62$\pm$0.05&   1.75&   0.67&          \\
   89&   J 135739.04-395820.9&13 57 39.04 -39 58 20.9& 21.45$\pm$0.11&  20.08$\pm$0.10&  19.24$\pm$0.08&   1.37&   0.84&          \\
   90&   J 135739.37-395637.4&13 57 39.37 -39 56 37.4&.............  &  20.51$\pm$0.15&  18.20$\pm$0.04&  ......&   2.31&         \\
   91&   J 135739.85-395411.5&13 57 39.85 -39 54 11.5& 19.33$\pm$0.03&  18.15$\pm$0.03&  17.44$\pm$0.03&   1.18&   0.71&          \\
   92&   J 135740.19-395527.8&13 57 40.19 -39 55 27.8& 22.38$\pm$0.29&  19.82$\pm$0.09&  18.59$\pm$0.05&   2.56&   1.23&          \\
   93&   J 135740.38-395441.6&13 57 40.38 -39 54 41.6& 14.52$\pm$0.01&  13.00$\pm$0.01&  12.45$\pm$0.01&   1.52&   0.55& 2        \\
   94&   J 135740.43-395512.9&13 57 40.43 -39 55 12.9& 20.92$\pm$0.12&  18.85$\pm$0.06&  17.88$\pm$0.03&   2.07&   0.97&          \\
   95&   J 135740.56-400007.7&13 57 40.56 -40 00 07.7& 19.85$\pm$0.04&  18.23$\pm$0.03&  17.66$\pm$0.02&   1.62&   0.57&          \\
   96&   J 135740.74-395453.6&13 57 40.74 -39 54 53.6& 18.99$\pm$0.02&  17.46$\pm$0.02&  16.91$\pm$0.01&   1.53&   0.55&          \\
   97&   J 135740.81-395705.4&13 57 40.81 -39 57 05.4& 19.88$\pm$0.04&  17.88$\pm$0.02&  16.87$\pm$0.01&   2.00&   1.01&          \\
   98&   J 135740.96-395831.9&13 57 40.96 -39 58 31.9& 18.09$\pm$0.01&  17.52$\pm$0.01&  17.16$\pm$0.02&   0.57&   0.36&          \\
   99&   J 135741.11-395801.5&13 57 41.11 -39 58 01.5& 17.37$\pm$0.01&  15.68$\pm$0.01&  14.92$\pm$0.01&   1.69&   0.76& 2        \\
  100&   J 135741.23-395444.9&13 57 41.23 -39 54 44.9& 19.62$\pm$0.03&  18.31$\pm$0.03&  17.59$\pm$0.02&   1.31&   0.72&          \\
  101&   J 135741.24-395747.5&13 57 41.24 -39 57 47.5& 19.55$\pm$0.03&  17.89$\pm$0.02&  17.01$\pm$0.01&   1.66&   0.88&          \\
  102&   J 135741.63-400038.1&13 57 41.63 -40 00 38.1& 17.09$\pm$0.01&  16.23$\pm$0.01&  15.95$\pm$0.01&   0.86&   0.28&          \\
  103&   J 135741.70-400003.4&13 57 41.70 -40 00 03.4& 20.51$\pm$0.08&  18.91$\pm$0.06&  17.99$\pm$0.03&   1.60&   0.92&          \\
  104&   J 135741.73-395522.8&13 57 41.73 -39 55 22.8& 20.52$\pm$0.07&  18.69$\pm$0.05&  17.65$\pm$0.02&   1.83&   1.04&          \\
  105&   J 135741.81-395658.2&13 57 41.81 -39 56 58.2& 21.21$\pm$0.10&  19.65$\pm$0.07&  18.71$\pm$0.05&   1.56&   0.94&          \\
  106&   J 135741.95-395630.3&13 57 41.95 -39 56 30.3& 21.92$\pm$0.18&  19.86$\pm$0.09&  18.66$\pm$0.05&   2.06&   1.20&          \\
  107&   J 135742.22-395757.5&13 57 42.22 -39 57 57.5& 19.40$\pm$0.02&  17.48$\pm$0.02&  16.69$\pm$0.01&   1.92&   0.79&          \\
  108&   J 135742.28-395846.7&13 57 42.28 -39 58 46.7& 18.13$\pm$0.01&  15.85$\pm$0.01&  14.38$\pm$0.01&   2.28&   1.47&  1       \\
  109&   J 135742.35-400047.2&13 57 42.35 -40 00 47.2& 20.13$\pm$0.05&  18.69$\pm$0.05&  18.65$\pm$0.05&   1.44&   0.04&          \\
  110&   J 135742.44-395818.7&13 57 42.44 -39 58 18.7& 20.56$\pm$0.06&  18.65$\pm$0.04&  17.72$\pm$0.02&   1.91&   0.93&          \\
  111&   J 135742.47-395458.2&13 57 42.47 -39 54 58.2& 20.10$\pm$0.05&  18.74$\pm$0.04&  18.02$\pm$0.03&   1.36&   0.72&          \\
  112&   J 135742.64-395628.0&13 57 42.64 -39 56 28.0& 22.23$\pm$0.29&  19.38$\pm$0.06&  18.37$\pm$0.04&   2.85&   1.01&          \\
  113&   J 135742.70-395506.3&13 57 42.70 -39 55 06.3& 20.42$\pm$0.07&  18.75$\pm$0.04&  18.04$\pm$0.03&   1.67&   0.71&          \\
  114&   J 135742.76-395640.1&13 57 42.76 -39 56 40.1& 20.64$\pm$0.07&  18.38$\pm$0.03&  17.39$\pm$0.02&   2.26&   0.99&          \\
  115&   J 135742.79-395525.0&13 57 42.79 -39 55 25.0& 21.43$\pm$0.10&  19.95$\pm$0.09&  18.95$\pm$0.06&   1.48&   1.00&          \\
  116&   J 135743.00-395654.3&13 57 43.00 -39 56 54.3& 20.63$\pm$0.07&  19.96$\pm$0.09&  19.23$\pm$0.07&   0.67&   0.73&          \\
  117&   J 135743.02-395953.3&13 57 43.02 -39 59 53.3& 18.94$\pm$0.02&  17.61$\pm$0.02&  17.12$\pm$0.02&   1.33&   0.49&          \\
  118&   J 135743.02-395443.3&13 57 43.02 -39 54 43.3& 20.16$\pm$0.06&  18.84$\pm$0.06&  18.14$\pm$0.04&   1.32&   0.70&          \\
  119&   J 135743.05-395500.4&13 57 43.05 -39 55 00.4& 20.78$\pm$0.08&  18.99$\pm$0.05&  18.47$\pm$0.04&   1.79&   0.52&          \\
  120&   J 135743.10-395624.0&13 57 43.10 -39 56 24.0& 16.60$\pm$0.01&  15.02$\pm$0.01&  14.18$\pm$0.01&   1.58&   0.84&  2       \\
  121&   J 135743.28-395714.4&13 57 43.28 -39 57 14.4& 21.86$\pm$0.26&  19.48$\pm$0.09&  18.31$\pm$0.04&   2.38&   1.17&          \\
  122&   J 135743.35-395801.6&13 57 43.35 -39 58 01.6& 21.19$\pm$0.19&  19.07$\pm$0.10&  18.61$\pm$0.05&   2.12&   0.46&          \\
  123&   J 135743.62-395745.8&13 57 43.62 -39 57 45.8& 19.59$\pm$0.03&  17.73$\pm$0.02&  16.87$\pm$0.01&   1.86&   0.86&          \\
  124&   J 135743.80-395825.5&13 57 43.80 -39 58 25.5& 18.85$\pm$0.02&  17.39$\pm$0.01&  16.76$\pm$0.01&   1.46&   0.63&          \\
  125&   J 135743.84-395922.0&13 57 43.84 -39 59 22.0& 22.41$\pm$0.23&  20.42$\pm$0.12&  18.70$\pm$0.06&   1.99&   1.72&          \\
  126&   J 135743.95-400014.8&13 57 43.95 -40 00 14.8& 17.38$\pm$0.01&  16.83$\pm$0.01&  16.40$\pm$0.01&   0.55&   0.43&          \\
  127&   J 135744.08-395953.1&13 57 44.08 -39 59 53.1& 19.78$\pm$0.04&  18.31$\pm$0.03&  17.62$\pm$0.02&   1.47&   0.69&          \\
  128&   J 135744.14-395737.7&13 57 44.14 -39 57 37.7& 20.24$\pm$0.06&  18.31$\pm$0.03&  17.41$\pm$0.02&   1.93&   0.90&          \\
  129&   J 135744.26-395716.8&13 57 44.26 -39 57 16.8& 21.42$\pm$0.14&  19.63$\pm$0.08&  18.89$\pm$0.06&   1.79&   0.74&          \\
  130&   J 135744.31-400014.5&13 57 44.31 -40 00 14.5& 20.34$\pm$0.06&  18.91$\pm$0.06&  18.45$\pm$0.04&   1.43&   0.46&          \\
  131&   J 135744.32-395853.5&13 57 44.32 -39 58 53.5& 20.27$\pm$0.50&  17.98$\pm$0.15&  16.69$\pm$0.15&   2.29&   1.29& 2        \\
  132&   J 135744.42-395650.4&13 57 44.42 -39 56 50.4& 11.26$\pm$0.10&  10.89$\pm$0.10&  10.64$\pm$0.10&   0.37&   0.25& 1, W77-6a\\
  133&   J 135744.50-395538.3&13 57 44.50 -39 55 38.3& 20.73$\pm$0.06&  19.89$\pm$0.09&  19.23$\pm$0.07&   0.84&   0.66&          \\
  134&   J 135744.53-395917.5&13 57 44.53 -39 59 17.5& 21.79$\pm$0.14&  20.39$\pm$0.12&  19.15$\pm$0.07&   1.40&   1.24&          \\
  135&   J 135744.63-395902.8&13 57 44.63 -39 59 02.8& 20.20$\pm$0.10&  19.31$\pm$0.11&  19.18$\pm$0.16&   0.89&   0.13&          \\
  136&   J 135744.63-395814.2&13 57 44.63 -39 58 14.2& 20.41$\pm$0.05&  19.10$\pm$0.05&  18.47$\pm$0.04&   1.31&   0.63&          \\
  137&   J 135744.66-395619.2&13 57 44.66 -39 56 19.2& 19.94$\pm$0.04&  18.65$\pm$0.04&  17.91$\pm$0.03&   1.29&   0.74&          \\
  138&   J 135744.69-395652.7&13 57 44.69 -39 56 52.7& 13.90$\pm$0.10&  12.68$\pm$0.10&  11.93$\pm$0.10&   1.22&   0.75& 1, W77-6b\\
  139&   J 135744.76-395556.6&13 57 44.76 -39 55 56.6& 20.45$\pm$0.05&  19.08$\pm$0.05&  18.57$\pm$0.05&   1.37&   0.51&          \\
  140&   J 135744.84-395644.5&13 57 44.84 -39 56 44.5& 19.42$\pm$0.03&  18.09$\pm$0.04&  17.56$\pm$0.04&   1.33&   0.53&          \\
  141&   J 135744.86-395946.7&13 57 44.86 -39 59 46.7& 17.46$\pm$0.01&  15.97$\pm$0.01&  15.31$\pm$0.01&   1.49&   0.66&          \\
  142&   J 135744.87-395619.6&13 57 44.87 -39 56 19.6& 20.08$\pm$0.05&  18.78$\pm$0.05&  18.32$\pm$0.05&   1.30&   0.46&          \\
  143&   J 135744.87-395730.9&13 57 44.87 -39 57 30.9& 18.72$\pm$0.01&  17.24$\pm$0.01&  16.58$\pm$0.01&   1.48&   0.66&          \\
  144&   J 135744.93-395751.4&13 57 44.93 -39 57 51.4& 20.66$\pm$0.07&  19.05$\pm$0.05&  18.47$\pm$0.04&   1.61&   0.58&          \\
  145&   J 135745.11-395550.0&13 57 45.11 -39 55 50.0& 20.64$\pm$0.06&  19.41$\pm$0.07&  18.98$\pm$0.06&   1.23&   0.43&          \\
  146&   J 135745.17-395801.2&13 57 45.17 -39 58 01.2& 19.11$\pm$0.02&  17.59$\pm$0.02&  16.99$\pm$0.01&   1.52&   0.60&          \\
  147&   J 135745.20-400001.3&13 57 45.20 -40 00 01.3& 20.52$\pm$0.06&  19.26$\pm$0.06&  18.55$\pm$0.04&   1.26&   0.71&          \\
  148&   J 135745.45-395932.4&13 57 45.45 -39 59 32.4& 17.92$\pm$0.01&  16.70$\pm$0.01&  16.22$\pm$0.01&   1.22&   0.48&          \\
  149&   J 135745.61-395921.7&13 57 45.61 -39 59 21.7& 20.53$\pm$0.06&  19.31$\pm$0.06&  18.70$\pm$0.06&   1.22&   0.61&          \\
  150&   J 135745.65-395656.0&13 57 45.65 -39 56 56.0& 17.65$\pm$0.01&  16.74$\pm$0.01&  16.43$\pm$0.01&   0.91&   0.31&          \\
  151&   J 135745.67-395621.5&13 57 45.67 -39 56 21.5& 20.22$\pm$0.04&  19.03$\pm$0.05&  18.78$\pm$0.06&   1.19&   0.25&          \\
  152&   J 135745.82-395847.3&13 57 45.82 -39 58 47.3& 20.28$\pm$0.07&  19.07$\pm$0.07&  18.56$\pm$0.07&   1.21&   0.51&          \\
  153&   J 135745.83-395727.1&13 57 45.83 -39 57 27.1& 20.38$\pm$0.05&  18.99$\pm$0.06&  18.36$\pm$0.04&   1.39&   0.63&          \\
  154&   J 135745.96-395930.7&13 57 45.96 -39 59 30.7& 20.52$\pm$0.07&  19.25$\pm$0.09&  18.95$\pm$0.07&   1.27&   0.30&          \\
  155&   J 135746.12-395649.8&13 57 46.12 -39 56 49.8& 20.38$\pm$0.09&  19.43$\pm$0.11&  19.56$\pm$0.09&   0.95&  -0.13&          \\
  156&   J 135746.31-395925.0&13 57 46.31 -39 59 25.0& 20.84$\pm$0.09&  19.64$\pm$0.09&  19.12$\pm$0.07&   1.20&   0.52&          \\
  157&   J 135746.44-395910.6&13 57 46.44 -39 59 10.6& 19.52$\pm$0.02&  18.53$\pm$0.03&  18.07$\pm$0.03&   0.99&   0.46&          \\
  158&   J 135746.46-395817.6&13 57 46.46 -39 58 17.6& 17.80$\pm$0.01&  16.62$\pm$0.01&  16.19$\pm$0.01&   1.18&   0.43&          \\
  159&   J 135746.48-395635.3&13 57 46.48 -39 56 35.3& 18.38$\pm$0.01&  17.27$\pm$0.01&  16.89$\pm$0.01&   1.11&   0.38&          \\
  160&   J 135746.53-395928.6&13 57 46.53 -39 59 28.6& 18.98$\pm$0.02&  17.87$\pm$0.02&  17.49$\pm$0.02&   1.11&   0.38&          \\
  161&   J 135746.62-395625.1&13 57 46.62 -39 56 25.1& 20.09$\pm$0.04&  19.03$\pm$0.05&  18.70$\pm$0.05&   1.06&   0.33&          \\
  162&   J 135746.78-395732.1&13 57 46.78 -39 57 32.1& 21.67$\pm$0.12&  20.38$\pm$0.13&  19.92$\pm$0.11&   1.29&   0.46&          \\
  163&   J 135746.80-395551.7&13 57 46.80 -39 55 51.7& 20.07$\pm$0.05&  18.85$\pm$0.06&  18.45$\pm$0.04&   1.22&   0.40&          \\
  164&   J 135746.88-395909.1&13 57 46.88 -39 59 09.1& 19.58$\pm$0.03&  18.67$\pm$0.04&  18.19$\pm$0.04&   0.91&   0.48&          \\
  165&   J 135746.90-395854.1&13 57 46.90 -39 58 54.1& 18.70$\pm$0.01&  17.59$\pm$0.02&  17.20$\pm$0.02&   1.11&   0.39&          \\
  166&   J 135746.98-395808.9&13 57 46.98 -39 58 08.9& 21.59$\pm$0.11&  20.38$\pm$0.12&  19.55$\pm$0.11&   1.21&   0.83&          \\
  167&   J 135747.01-395947.1&13 57 47.01 -39 59 47.1& 19.56$\pm$0.03&  18.65$\pm$0.04&  18.09$\pm$0.03&   0.91&   0.56&          \\
  168&   J 135747.03-395554.6&13 57 47.03 -39 55 54.6& 18.09$\pm$0.01&  17.10$\pm$0.01&  16.79$\pm$0.01&   0.99&   0.31&          \\
  169&   J 135747.53-395629.9&13 57 47.53 -39 56 29.9& 16.47$\pm$0.01&  15.79$\pm$0.01&  15.60$\pm$0.01&   0.68&   0.19&          \\
  170&   J 135747.52-395924.1&13 57 47.52 -39 59 24.1& 20.28$\pm$0.04&  19.43$\pm$0.08&  18.92$\pm$0.06&   0.85&   0.51&          \\
  171&   J 135747.91-395911.2&13 57 47.91 -39 59 11.2& 20.24$\pm$0.05&  19.02$\pm$0.06&  19.05$\pm$0.07&   1.22&  -0.03&          \\
  172&   J 135747.94-395840.6&13 57 47.94 -39 58 40.6& 19.75$\pm$0.03&  18.76$\pm$0.04&  18.23$\pm$0.04&   0.99&   0.53&          \\
  173&   J 135748.01-395706.3&13 57 48.01 -39 57 06.3& 21.15$\pm$0.09&  19.79$\pm$0.08&  18.96$\pm$0.07&   1.36&   0.83&          \\
  174&   J 135748.03-395847.0&13 57 48.03 -39 58 47.0& 12.76$\pm$0.01&  11.89$\pm$0.01&  11.61$\pm$0.01&   0.87&   0.28&          \\
  175&   J 135748.01-395652.3&13 57 48.01 -39 56 52.3& 16.37$\pm$0.01&  15.42$\pm$0.01&  15.05$\pm$0.01&   0.95&   0.37&          \\
  176&   J 135748.28-395655.9&13 57 48.28 -39 56 55.9& 20.64$\pm$0.08&  19.60$\pm$0.08&  19.03$\pm$0.08&   1.04&   0.57&          \\
  177&   J 135748.51-395945.2&13 57 48.51 -39 59 45.2& 19.50$\pm$0.03&  18.29$\pm$0.04&  18.15$\pm$0.04&   1.21&   0.14&          \\
  178&   J 135748.72-395620.5&13 57 48.72 -39 56 20.5& 18.23$\pm$0.01&  17.23$\pm$0.01&  16.95$\pm$0.01&   1.00&   0.28&          \\
  179&   J 135748.80-395802.1&13 57 48.80 -39 58 02.1& 20.75$\pm$0.06&  19.68$\pm$0.07&  19.02$\pm$0.08&   1.07&   0.66&          \\
  180&   J 135748.81-395729.5&13 57 48.81 -39 57 29.5& 19.69$\pm$0.03&  18.69$\pm$0.04&  18.16$\pm$0.04&   1.00&   0.53&          \\
  181&   J 135748.88-395842.6&13 57 48.88 -39 58 42.6& 19.03$\pm$0.02&  18.04$\pm$0.02&  17.67$\pm$0.02&   0.99&   0.37&          \\
  182&   J 135748.89-395706.5&13 57 48.89 -39 57 06.5& 20.04$\pm$0.05&  19.00$\pm$0.05&  18.45$\pm$0.06&   1.04&   0.55&          \\
  183&   J 135749.06-395618.5&13 57 49.06 -39 56 18.5& 18.83$\pm$0.01&  17.94$\pm$0.02&  17.57$\pm$0.02&   0.89&   0.37&          \\
  184&   J 135749.14-395905.6&13 57 49.14 -39 59 05.6& 18.91$\pm$0.02&  18.21$\pm$0.03&  18.10$\pm$0.03&   0.70&   0.11&          \\
  185&   J 135749.47-395623.1&13 57 49.47 -39 56 23.1& 20.35$\pm$0.05&  19.42$\pm$0.06&  18.55$\pm$0.08&   0.93&   0.87&          \\
  186&   J 135749.69-395638.9&13 57 49.69 -39 56 38.9& 20.82$\pm$0.07&  19.84$\pm$0.09&  19.11$\pm$0.10&   0.98&   0.73&          \\
  187&   J 135749.77-395802.3&13 57 49.77 -39 58 02.3& 19.71$\pm$0.03&  18.41$\pm$0.03&  17.98$\pm$0.03&   1.30&   0.43&          \\
  188&   J 135750.03-395720.6&13 57 50.03 -39 57 20.6& 20.21$\pm$0.05&  19.28$\pm$0.07&  18.77$\pm$0.07&   0.93&   0.51&          \\
  189&   J 135750.04-395635.1&13 57 50.04 -39 56 35.1& 20.69$\pm$0.05&  19.82$\pm$0.08&  19.12$\pm$0.07&   0.87&   0.70&          \\
  190&   J 135750.06-395728.4&13 57 50.06 -39 57 28.4& 21.17$\pm$0.10&  19.93$\pm$0.11&  19.71$\pm$0.12&   1.24&   0.22&          \\
  191&   J 135750.10-395753.2&13 57 50.10 -39 57 53.2& 20.69$\pm$0.06&  19.54$\pm$0.07&  18.94$\pm$0.06&   1.15&   0.60&          \\
  192&   J 135750.11-395909.5&13 57 50.11 -39 59 09.5& 18.52$\pm$0.01&  17.80$\pm$0.02&  17.58$\pm$0.02&   0.72&   0.22&          \\
  193&   J 135750.18-395742.4&13 57 50.18 -39 57 42.4& 19.18$\pm$0.02&  18.18$\pm$0.03&  17.82$\pm$0.03&   1.00&   0.36&          \\
  194&   J 135750.20-395629.1&13 57 50.20 -39 56 29.1& 19.30$\pm$0.03&  18.53$\pm$0.04&  18.12$\pm$0.04&   0.77&   0.41&          \\
  195&   J 135750.26-395901.8&13 57 50.26 -39 59 01.8& 19.68$\pm$0.03&  18.83$\pm$0.04&  18.58$\pm$0.05&   0.85&   0.25&          \\
  196&   J 135750.39-395650.7&13 57 50.39 -39 56 50.7& 16.92$\pm$0.01&  16.04$\pm$0.01&  15.68$\pm$0.01&   0.88&   0.36&          \\
  197&   J 135750.56-395721.6&13 57 50.56 -39 57 21.6& 21.01$\pm$0.12&  19.71$\pm$0.12&  19.32$\pm$0.13&   1.30&   0.39&          \\
  198&   J 135750.60-395757.0&13 57 50.60 -39 57 57.0& 17.11$\pm$0.01&  16.08$\pm$0.01&  15.68$\pm$0.01&   1.03&   0.40&          \\
  199&   J 135750.65-395905.2&13 57 50.65 -39 59 05.2& 18.07$\pm$0.01&  16.97$\pm$0.01&  16.63$\pm$0.01&   1.10&   0.34&          \\
  200&   J 135750.79-395903.8&13 57 50.79 -39 59 03.8& 15.16$\pm$0.01&  14.55$\pm$0.01&  14.38$\pm$0.01&   0.61&   0.17&          \\
  201&   J 135750.84-395610.8&13 57 50.84 -39 56 10.8& 16.23$\pm$0.01&  15.52$\pm$0.01&  15.31$\pm$0.01&   0.71&   0.21&          \\
  202&   J 135751.06-395845.1&13 57 51.06 -39 58 45.1& 19.39$\pm$0.02&  18.51$\pm$0.03&  18.08$\pm$0.03&   0.88&   0.43&          \\
  203&   J 135751.11-395909.9&13 57 51.11 -39 59 09.9& 17.20$\pm$0.01&  16.60$\pm$0.01&  16.45$\pm$0.01&   0.60&   0.15&          \\
  204&   J 135751.13-395851.5&13 57 51.13 -39 58 51.5& 16.08$\pm$0.01&  15.09$\pm$0.01&  14.88$\pm$0.01&   0.99&   0.21&          \\
  205&   J 135751.26-395617.9&13 57 51.26 -39 56 17.9& 20.08$\pm$0.04&  19.31$\pm$0.06&  19.01$\pm$0.06&   0.77&   0.30&          \\
  206&   J 135751.28-395725.0&13 57 51.28 -39 57 25.0& 20.31$\pm$0.07&  19.29$\pm$0.08&  18.59$\pm$0.06&   1.02&   0.70&          \\
  207&   J 135751.40-395850.3&13 57 51.40 -39 58 50.3& 16.10$\pm$0.01&  15.09$\pm$0.01&  14.81$\pm$0.01&   1.01&   0.28&          \\
  208&   J 135751.54-395733.3&13 57 51.54 -39 57 33.3& 20.01$\pm$0.04&  19.07$\pm$0.05&  18.43$\pm$0.05&   0.94&   0.64&          \\
  209&   J 135751.65-395805.0&13 57 51.65 -39 58 05.0& 20.88$\pm$0.08&  19.56$\pm$0.10&  19.24$\pm$0.09&   1.32&   0.32&          \\
  210&   J 135752.61-395816.5&13 57 52.61 -39 58 16.5& 18.31$\pm$0.01&  17.02$\pm$0.01&  16.59$\pm$0.01&   1.29&   0.43&          \\
  211&   J 135752.78-395815.3&13 57 52.78 -39 58 15.3& 18.42$\pm$0.01&  17.50$\pm$0.01&  17.10$\pm$0.02&   0.92&   0.40&          \\
  212&   J 135752.99-395843.4&13 57 52.99 -39 58 43.4& 15.91$\pm$0.01&  15.21$\pm$0.01&  15.02$\pm$0.01&   0.70&   0.19&          \\
  213&   J 135753.05-395654.2&13 57 53.05 -39 56 54.2& 20.88$\pm$0.09&  19.91$\pm$0.12&  19.26$\pm$0.11&   0.97&   0.65&          \\
  214&   J 135753.14-395729.2&13 57 53.14 -39 57 29.2& 20.30$\pm$0.05&  19.54$\pm$0.07&  18.93$\pm$0.07&   0.76&   0.61&          \\
  215&   J 135753.55-395816.0&13 57 53.55 -39 58 16.0& 20.41$\pm$0.05&  19.16$\pm$0.06&  18.81$\pm$0.07&   1.25&   0.35&          \\
  216&   J 135753.74-395730.8&13 57 53.74 -39 57 30.8& 15.68$\pm$0.01&  15.05$\pm$0.01&  14.76$\pm$0.01&   0.63&   0.29& 1, I\_68 \\
  217&   J 135753.90-395840.9&13 57 53.90 -39 58 40.9& 16.08$\pm$0.01&  15.05$\pm$0.01&  14.78$\pm$0.01&   1.03&   0.27&          \\
  218&   J 135754.80-395804.4&13 57 54.80 -39 58 04.4& 16.52$\pm$0.01&  15.59$\pm$0.01&  15.34$\pm$0.01&   0.93&   0.25&          \\
  219&   J 135755.04-395742.9&13 57 55.04 -39 57 42.9& 16.23$\pm$0.01&  15.54$\pm$0.01&  15.33$\pm$0.01&   0.69&   0.21&          \\
  220&   J 135755.06-395702.2&13 57 55.06 -39 57 02.2& 18.90$\pm$0.02&  17.91$\pm$0.02&  17.79$\pm$0.03&   0.99&   0.12&          \\
  221&   J 135755.45-395727.7&13 57 55.45 -39 57 27.7& 18.78$\pm$0.02&  17.87$\pm$0.02&  17.59$\pm$0.02&   0.91&   0.28&          \\
  222&   J 135755.64-395759.2&13 57 55.64 -39 57 59.2& 20.41$\pm$0.06&  19.48$\pm$0.07&  19.01$\pm$0.08&   0.93&   0.47&          \\
  223&   J 135755.84-395821.3&13 57 55.84 -39 58 21.3& 15.22$\pm$0.01&  14.38$\pm$0.01&  14.10$\pm$0.01&   0.84&   0.28&          \\
  224&   J 135756.78-395729.1&13 57 56.78 -39 57 29.1& 18.66$\pm$0.01&  18.00$\pm$0.03&  17.86$\pm$0.03&   0.66&   0.14&          \\
  225&   J 135756.83-395744.7&13 57 56.83 -39 57 44.7& 18.81$\pm$0.02&  18.10$\pm$0.03&  17.80$\pm$0.03&   0.71&   0.30&          \\
  226&   J 135756.89-395736.1&13 57 56.89 -39 57 36.1& 17.94$\pm$0.01&  17.09$\pm$0.01&  16.72$\pm$0.01&   0.85&   0.37&          \\
  227&   J 135757.52-395746.4&13 57 57.52 -39 57 46.4& 15.44$\pm$0.01&  14.87$\pm$0.01&  14.73$\pm$0.01&   0.57&   0.14&         \\
\end{longtable}  
Comments: 1: CG 12 cluster member \quad \quad 2: Potential CG 12 cluster member
}

\end{document}